\documentclass{aa}  

\usepackage{graphicx}
\usepackage{txfonts}
\begin{document}

   \title{Ca Triplet Metallicities and Velocities for  twelve Globular Clusters towards the Galactic Bulge}
 
 \titlerunning{Ca Triplet Metallicities and Velocities for twelve Globular Clusters towards the Bulge}  

   \author{D. Geisler\inst{1,2,3}
          \and
          M.C. Parisi\inst{4,5}
          \and 
          B. Dias\inst{6}
          \and
          S. Villanova\inst{1} 
          \and 
          F. Mauro\inst{7} 
          \and 
          I. Saviane\inst{8} 
          \and 
          R.E. Cohen\inst{9} 
          \and 
          C. Moni Bidin\inst{7}
          \and 
          D. Minniti\inst{10,11}
          }

\institute{
    Departmento de Astronomía, Universidad de Concepción, Casilla 160-C, Concepción, Chile
    \and 
    Departamento de Física y Astronomía, Universidad de La Serena, Avenida Juan Cisternas 1200, La Serena, Chile
    \and
    Instituto de Investigación Multidisciplinario en Ciencia y Tecnología, Universidad de La Serena Benavente 980, La Serena, Chile
    \\
    \email{dgeisler@astroudec.cl}
    \and
    %3
    Observatorio Astron\'omico, Universidad Nacional de C\'ordoba, Laprida 854, X5000BGR, C\'ordoba, Argentina.
         \and
    %4
    Instituto de Astronom{\'\i}a Te\'orica y Experimental (CONICET-UNC), Laprida 854, X5000BGR, C\'ordoba, Argentina.
  	 %1
         \and
    Instituto de Alta Investigaci\'on, Sede Esmeralda, Universidad de Tarapac\'a, Av. Luis Emilio Recabarren 2477, Iquique, Chile%\\
             \and
    %5
    Instituto de Astronom{\'\i}a, Universidad Cat\'olica del Norte, Av. Angamos 0610, Antofagasta, Chile.
    \and
        %7
    European Southern Observatory, Casilla 19001, Santiago, Chile
    \and
    %6
    Department of Physics and Astronomy, Rutgers the State University of New Jersey, 136 Frelinghuysen Road., Piscataway, NJ 08854, USA
    \and
    %8
    Instituto de Astrof{\'\i}sica, Facultad de Ciencias Exactas, Universidad Andr\'es Bello, Av. Fern\'andez Concha 700, Las Condes, Santiago, Chile
    \and
    %9 
    Vatican Observatory, V00120 Vatican City State, Italy
\\
   }

   \date{Received; accepted ...}

% \abstract{}{}{}{}{} 
% 5 {} token are mandatory
 
  \abstract
   {Globular clusters (GCs) are excellent tracers of the formation and early evolution of the Milky Way. 
The bulge GCs (BGCs) are particularly important because they can reveal vital information about the oldest,   
in-situ component of the Milky Way.
}
   {We aim at deriving mean  metallicities and radial velocities for 13 GCs that lie towards the bulge and are generally associated with this component. This region is observationally challenging because of high extinction and stellar density, hampering optical studies of these and similar BGCs, making most previous determinations of these parameters quite uncertain. }
   {We use near infrared  low resolution spectroscopy with the FORS2 instrument on the VLT to measure the wavelengths and equivalent widths of the CaII triplet (CaT) lines for a number of stars per cluster. We derive radial velocities, ascertain membership and apply known calibrations to determine  metallicities for cluster members, for a mean of 11 members per cluster.
   Unfortunately, one of our targets, VVV-GC002, which is the closest GC to the Galactic center, 
   %was observed long before Gaia results and 
   turned out not to have any members in our sample.}
   {We derive mean cluster RV values to 3 km/s, and mean metallicities to 0.05 dex.
   We find generally good agreement with previous determinations for both metallicity and velocity. On average, our metallicities are 0.07 dex more metal-rich than Harris (2010), with a standard deviation of the difference of 0.25 dex.  
   Our sample has metallicities lying between -0.21 and -1.64 and  is  %rather uniformly 
   distributed between the traditional metal-rich BGC peak near [Fe/H]$\sim$-0.5 and a more metal-poor peak around [Fe/H]$\sim$-1.1, which  has recently been identified.
   These latter are candidates for the oldest GCs in the Galaxy, if blue horizontal branches are present, and include BH\,261, NGC\,6401, NGC\,6540, NGC\,6642, and Terzan\,9. Finally, Terzan 10 is even more metal-poor. However, dynamically, Terzan 10 is likely an intruder from the halo, possibly associated with the Gaia-Enceladus or Kraken accretion events. Terzan\,10 is also confirmed as an Oosterhoff type II GC based on our results.}
   {The CaT technique is an excellent method for deriving mean metallicities and velocities for heavily obscured GCs. Our sample provides reliable mean values for both of these key properties for an important sample of previously poorly-studied BGCs from spectroscopy of a significant number of members per cluster. 
   We emphasize that the more metal-poor GCs are excellent candidates for being ancient relics of bulge formation. The lone halo intruder in our sample, Terzan 10, is conspicuous for also having by far the lowest metallicity, and casts doubt on the possibility of any bonafide BGCs at metallicities below about $\sim$ -1.5.}


\maketitle
%%%%%%%%%%%%%%%%%%%%%%%%%%%%%%%%%%%%%%
%___________________________________________________________
\section{Introduction}
\label{sec:int}

The formation and evolution of the Milky Way bulge has long been of salient astrophysical interest, both in the context of the Milky Way itself as well as how our bulge relates to similar structures in other galaxies \citep[see e.g.][and references therein]{gonzalez+16,barbuy+18,saviane+20}. We now believe that our bulge formed via several processes.
On the one hand, a pressure-supported component formed in situ at the beginning of the Milky Way's assembly, which is the classical, spheroidal bulge, containing a small fraction of the bulge total mass ($\sim 1\%$ as indicated by old and metal-poor RR Lyrae, for example).
Later, the boxy-peanut or X-shaped bulge/bar was formed outside-in from inner disc instabilities, containing most of the bulge mass ($\sim 90\%$ as indicated by red clump giant stars), leading to a present-day bulge displaying evidence of both processes (see e.g. \citealp{babusiaux+10}, \citealp{zoccali+16}, \citealp{barbuy+18} and references therein).

This evidence comes from a variety of tracers. RR Lyrae stars represent the old stellar populations of the bulge and display a spheroidal distribution. Additionally, their metallicities peak around [Fe/H]=-1.0 (\citealp{dekany+13}, but see also \citealp{pietrukowics+15}). Red clump stars, with a wider range of age sensitivity, show a bimodal bulge, with two main components: metal-poor stars are more concentrated in a spherical shape and with slower rotation, whereas metal-rich stars are distributed in a boxy shape with a faster rotation \citep[e.g.][]{kunder+16,zoccali+17}. Lastly, GCs are also excellent tracers of the oldest stellar populations in the Galactic bulge. True BGCs most likely formed in situ well before the formation of the bar (Bovy et al. 2019) and stayed confined within the bulge. By the time the bar buckled into a boxy-peanut shape, it trapped the existing BGCs of all metallicities within the inner bulge \citep{rossi+15,bica+16}. Their orbits, and likely also metallicity, distinguish them from thick disk or inner halo GC intruders \citep{perez-villegas+20}. Therefore, it is important to analyse the metallicity and kinematics of the bonafide  BGCs to provide further constraints on the formation and extension of the classical spheroidal component of the Galactic bulge.

\begin{figure*}
    \centering
    \includegraphics[width=\textwidth]{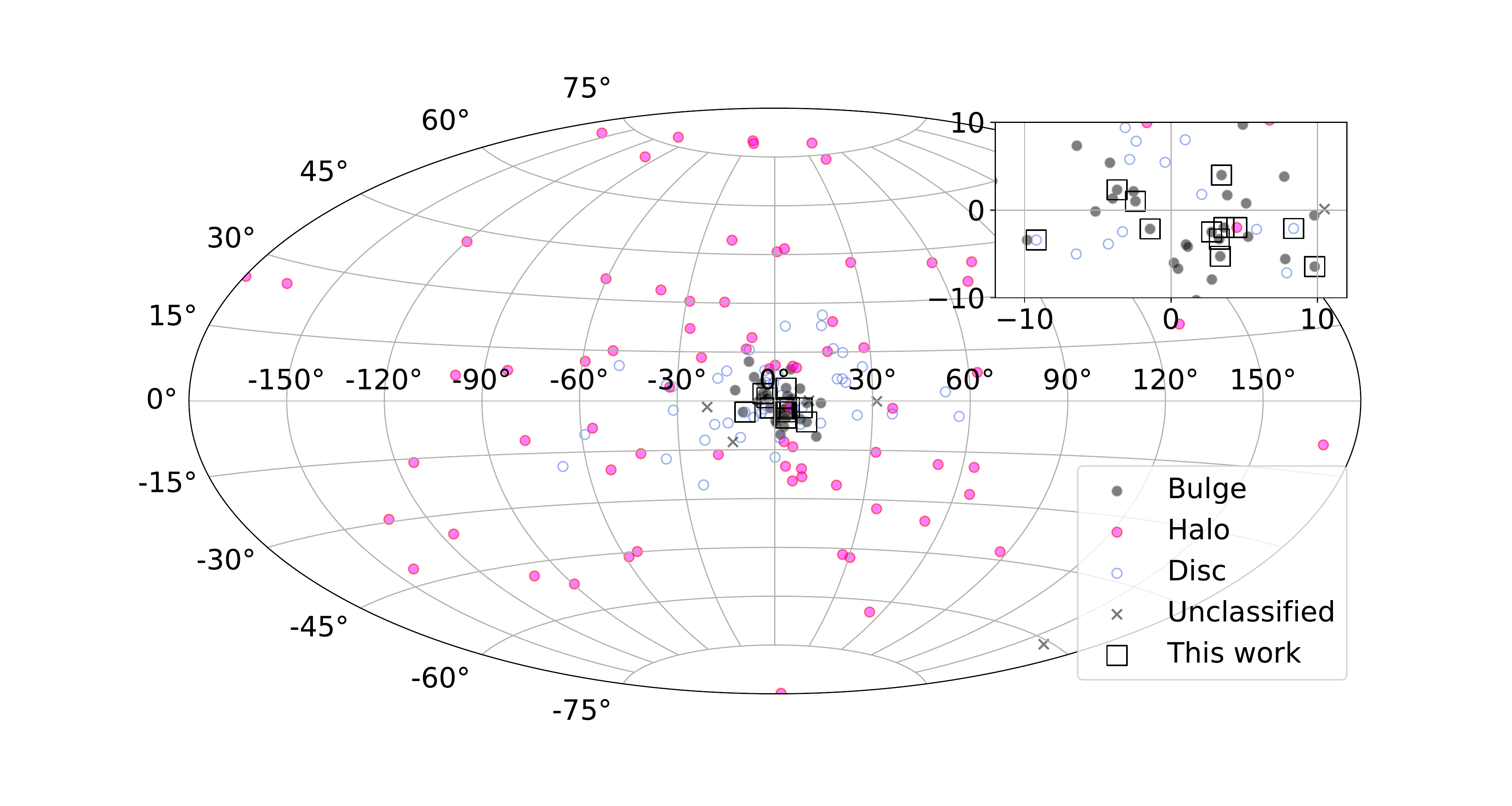}
    \caption{Aitoff distribution in Galactic coordinates of all Milky Way GCs from the Harris catalogue. The classification of Bulge, Disc, Halo comes from \citet{dias+16} updated with classification by \citet{perez-villegas+20} when available and are indicated by circles in different colours. The GCs studied in this work are highlighted with empty squares. The inset shows a zoomed in region around the GCs analysed in this work. Assuming this classification, nine GCs belong to the bulge, two to the disc and one to the halo.}
    \label{fig:aitoff}
\end{figure*}

It has been known for decades that Galactic GCs can be separated into two populations in terms of their metallicity and spatial distributions: a more metal-poor ([Fe/H]$\sim $-1.6) halo component, and a more metal-rich ([Fe/H]$\sim$ -0.6), centrally concentrated disc/bulge component \citep[]{zinn85,minniti95,dias+16}.
Unfortunately, detailed observations of this latter component have been severely limited due to extinction, especially in the optical. Nevertheless, with the advent of  infrared detectors and dedicated surveys like the 
Vista Variables in the Via Lactea (VVV) \citep{Minniti+10} and Apache Point Observatory Galactic Evolution Experiment (APOGEE) \citep{majewski+17},  it is now possible to observe in much greater detail GCs located towards the bulge. Such observations have suggested an additional sub-population of bulge GCs,  with metallicities substantially below that of the traditional bulge population,
with a peak around [Fe/H]$\sim$ -1.1 \citep{barbuy+06,barbuy+09,bica+16,barbuy+18proc} but sharing similar chemical and dynamical patterns \citep[][fig.12]{barbuy+18_6558}. 
Indeed, there may even be yet another population of BGCs with even lower metallicities. \citet{perez-villegas+20} find a small peak around [Fe/H]$\sim$ -1.5.
Many of these intermediate and lower metallicity BGCs also have a blue horizontal branch, which makes them excellent candidates for the oldest GCs \citep{lee+94,dias+16} in the Milky Way, with ages approaching the age of the Universe \citep[][]{kerber+19}.
Indeed, without invoking HB models, it was recently observationally confirmed that BGCs with blue HBs for their metallicity are quite old, with remarkable consistency: \citet{cohen+21} found a mean age of 12.9$\pm 0.4$ Gyr for eight BGCs. 

Interestingly, the latest study of bulge field stars, which now includes high resolution, high S/N spectra of many thousands of genuine bulge stars, also reveals a trimodal metallicity distribution \citep{rojas-Arriagada+20}.
However, the peaks are significantly offset from those of the BGCs, with means of +0.32, -0.17 and -0.66. 
Clearly, it is important to enhance the number of BGCs with  accurate metallicities as well as velocities, both of which are only poorly known in general. Besides deriving these key parameters for these clusters to improve our limited knowledge of them, 
this will help us derive a definitive BGC metallicity distribution to compare to its field star counterpart, select a larger population of GCs of relatively low metallicity which are the best relics to explore the ancient bulge component, investigate the origin of BGCs by determining their orbits by combining accurate radial velocities with the exquisite Gaia proper motions,
and identify possible halo interlopers within the BGC census.

As noted, extinction is a big challenge for optical high-resolution spectroscopy of BGCs, which is traditionally the best source of accurate metallicities. The advent of the APOGEE-2 main survey and the  complementary CAPOS (bulge Cluster APOgee Survey)  \citep{geisler+21} projects have gone a long way to 
help alleviate our previous lack of knowledge of key parameters for a number of BGCs,  
using near-infrared high-resolution spectroscopy to derive metallicities, chemical abundances and velocities. Another successful alternative is using low-resolution optical spectroscopy that can reach higher signal-to-noise than high-resolution spectroscopy using similar or shorter exposure times. \cite{dias+16} increased significantly the number of bulge GCs with known spectroscopic metallicities, in particular extending and superseding the previously adopted metallicity scale for GCs \citep{carretta+09} by adding metal-rich GCs to the homogeneous sample. 

A further complementary technique uses the near-infrared CaII triplet (CaT) lines as metallicity indicators. This is a very efficient way to build up a large sample of accurate metallicity and velocity measurements in BGCs. The CaT technique has many advantages. The brightest stars in clusters older than $\sim$1 Gyr are the red giants, and are thus the
natural targets for precision measurements of cluster abundances and velocities. The CaT lines are extremely strong and near the peak 
flux of unreddened RGB stars, and the technique only requires low resolution
(R $\sim$3000). Because there are many giants in a typical GC, the derived mean abundance can be made much more robust than that based on only one or a few stars, taking advantage of a multiplexing spectrograph. A reasonable sample of stars must also be observed in order
to ensure sufficient cluster members, especially in BGCs where membership on the bright RGB may be as low as 20 \% due to field contamination \citep{saviane+12} (for observations taken prior to Gaia results, such as these). This technique can derive metallicities even in the most extincted areas of the Galactic bulge. Many authors have confirmed the accuracy and repeatability of CaT abundance measurements in combination
with broad-band photometry and shown its very high sensitivity to metallicity and insensitivity to age \citep[e.g.,][]{cole+04}.

In view of all of these advantages, many BGCs have now had a sample of their RGB stars observed using CaT \citep{rutledge+97a,saviane+12,mauro+14,vasquez+18}. However, when we began this study, only about half of the known sample of BGCs had been observed, and we successfully proposed to investigate the remaining sample, with the main goal of completing the sample, essentially doubling the number of bulge GCs with metallicities and velocities from CaT. In particular, the present sample also includes VVV GC002, Terzan 1, Terzan 2,  Terzan 6,  Terzan 9, Djorg 2, NGC 6401 and NGC6642, GCs that are deemed to have the closest perigalactica to the Galactic center \citep{minniti+21}.

This paper is organised as follows. In Section 2 we present our cluster sample and describe our target selection. Section 3 discusses the observations and reduction procedure. The measurement of velocities and equivalent widths using the CaT lines is given in Section 4. Section 5 describes the membership selection and metallicity derivation. We compare our results with previous literature values in Section 6.
Section 7  discusses the nature of our sample and the bulge metallicity distribution, and we close our paper with the main conclusions.

%%%%%%%%%%%%%%%%%%%%%%%%%%%%%%%%%%%%%%
%___________________________________________________________
\section{Cluster sample and target selection}
\label{sec:sample}

Our original targets included
all GCs appearing within the VVV survey which had not yet had spectra of individual stars obtained at the time of our observing
(begun around June 2012). These objects were obvious choices, given their central location in the bulge and the fact that the VVV photometry provided everything required to carryout a successful spectroscopic program, including CMD and radial profile information used to select stars, the astrometry needed to position the slits (thus obviating the need for large-overhead pre-images), as well as the photometry used to calibrate the CaT technique. 
A total of 17 GCs were targeted, including principally GCs from the catalog of \citet{Harris1996} (version 2010, \citealt[]{harris10} - hereafter H10)  but also two GCs discovered early during the course of the VVV survey: VVV CL001 \citep{minniti+11} and VVV CL002 \citep{moni-bidin+11}. Both of these clusters are particularly interesting, as VVV CL001 appears to be the most metal-poor surviving GC in the inner Galaxy \citep{Fernandez-Trincado+21} and VVV CL002 is the closest known GC to the Galactic center \citep{minniti+21}.

During our first allocation, spectra for only 4 GCs of our sample were obtained,
and we subsequently successfully proposed to finish our program the following year. However, in the interim, we became aware that a competing program had obtained data for several of our original targets, which we then eliminated from our list. In addition, data for one of our original targets, 2MS-GC02, had very low S/N, despite having the longest integration, and we also eliminated this cluster. Data for the remaining 14 GCs observed during both runs were obtained as described below. Results for one of our sample, VVV CL001, are given in a companion paper and this cluster will not be discussed further here.
The GCs analysed in this work are identified in Figure \ref{fig:aitoff}. We emphasize that the main reason that our clusters have received little attention before is that they are among the most reddened GCs in the Galaxy, with E(B-V) estimates ranging up to almost 3.

\begin{figure}
    \centering
    \includegraphics[width=\columnwidth]{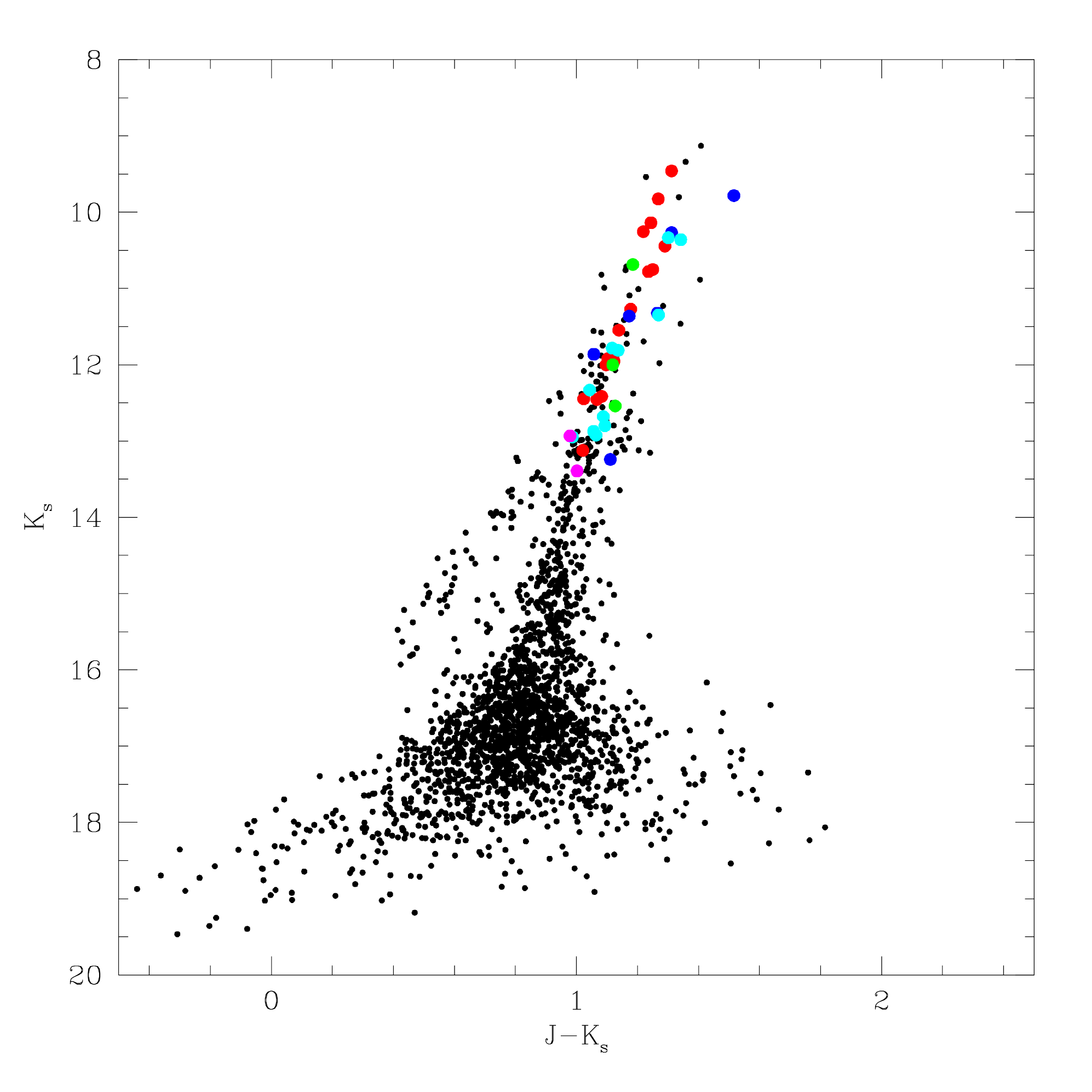}
    \caption{Color Magnitude Diagram of the cluster NGC 6401. Large symbols represent our spectroscopic targets colored according to our membership classification.
    Blue symbols represent non-members located beyond the adopted cluster radius. Cyan, green and pink symbols are stars discarded because they have incompatible radial velocities, metallicities or proper motions, respectively. Red points are the stars adopted as final cluster members.}
    \label{f:N6401_cmd}
\end{figure}

The individual spectroscopic targets are red giant stars selected from the cluster VVV CMDs (see Figure \ref{f:N6401_cmd}).
Bright, relatively isolated stars lying along the principal RGB and close to the cluster center were prioritized.
As remarked above, our observations long preceded Gaia data, so unfortunately no proper motions were available to help select members at this stage.

%%%%%%%%%%%%%%%%%%%%%%%%%%%%%%%%%%%%%%
%___________________________________________________________
\section{Observations and reduction}
\label{sec:observations}

Using the FORS2 instrument \citep{appenzeller+98} on the Very Large Telescope (Paranal, Chile), we obtained spectra of $\sim$ 540 red giant stars.
Observations were performed as part of the programs 089.D-0392 and 091.D-0389 (D. Geisler PI) in service mode. We used FORS2 in mask exchange unit mode,
with the 1028z+29 grism and OG590+32 filter.   FORS2 has two CCDs (2000 $\times$ 4000 pixels each detector): the master and the secondary chips, which have
a readout noise of 2.9 and 3.15 electrons, respectively, and a gain of 0.7 e${^-}$ ADU. In most cases, the cluster was observed on the master CCD, while the secondary detector was used for observations of field stars. In a few exceptions (Terzan 9 and Terzan 12), the cluster occupied part of both CCDs. We located between
33 and 66 slits in each total frame (master $+$ secondary CCDs), 1'' wide and 4 $-$ 8'' long. Pixels were binned 2$\times$2, yielding a
plate scale of 0.25'' pixel$^{-1}$, and a dispersion of $\sim$ 0.85 \AA \space pixel$^{-1}$. Resulting spectra cover a range of 1750  \AA $\rm$ (7750 $-$ 9500 \AA),
with a central wavelength of 8600 \AA, coincident with the region of the CaT lines. 
Relevant information is given in Table \ref{t:info}, where we include the cluster ID, equatorial coordinates, 
the $K_s$ magnitude of
the red horizontal branch, derived from the VVV data (see below) and the reddening. 

\begin{table*}
\caption{Observed Bulge Globular Clusters }             % title of Table
\label{t:info}      % is used to refer this table in the text
\centering                          % used for centering table
\footnotesize
\begin{tabular}{lccccccc}
\hline
\multicolumn{1}{c}{Cluster} & 
\multicolumn{1}{c}{Alternative} & 

\multicolumn{1}{c}{R.A.} & 
\multicolumn{1}{c}{Dec} & 
\multicolumn{1}{c}{$K_{s,HB}$} &
\multicolumn{1}{c}{E(B-V)$^{*}$} \\
\multicolumn{1}{c}{} & 
\multicolumn{1}{c}{designation} &
\multicolumn{1}{c}{J2000}  & 
\multicolumn{1}{c}{J2000}  & 
\multicolumn{1}{c}{mag} &
\multicolumn{1}{c}{} \\
\hline
\noalign{\smallskip}
BH 261    & AL 3             & 18 14 06.6 & -28 38 06 &  12.85 $\pm$ 0.10 & 0.36   \\
          & ESO 456-78       &            &           &   &   &          \\
          & MWSC 2847        &            &           &   &   &          \\
\hline
Djorg 2   & ESO 456-38       & 18 01 49.1 & -27 49 33 &  12.92 $\pm$ 0.10 & 0.94 \\
          & MWSC 2779        &            &           &   &   &         \\
\hline
NGC 6401  & ESO 521-11       & 17 38 36.6 & -23 54 34 &  13.17 $\pm$ 0.05 & 0.72  \\
          & MWSC 2653        &            &           &     &  &             \\
\hline
NGC 6540  & Djorg 3          & 18 06 08.6 & -27 45 55 &  12.64 $\pm$ 0.05 & 0.66 \\
          & BH 258           &            &           &     &  &             \\
          & MWSC 2804        &            &           &     &  &             \\
\hline
NGC 6642  & ESO 522-32       & 18 31 54.1 & -23 28 31 &  13.14 $\pm$ 0.04 & 0.40 \\
          & MWSC 2941        &            &           &     &  &             \\
\hline
Terzan 1  & ESO 455-23       & 17 35 47.2 & -30 28 54 &  13.45 $\pm$ 0.10 & 1.99   \\
          & Haute-Provence 2 &            &           &     &  &           \\
          & BH 235           &            &           &     &  &           \\
          & MWSC 2635        &            &           &     &  &           \\
\hline
Terzan 2  & ESO 454-29       & 17 27 33.1 & -30 48 08 &  13.70 $\pm$ 0.05 & 1.87 \\
          & Haute-Provence 3 &            &           &     &  &             \\
          & BH 228           &            &           &     &  &             \\
          & MWSC 2600        &            &           &     &  &              \\
\hline
Terzan 6  & ESO 455-49       & 17 50 46.4 & -31 16 31 &  13.80 $\pm$ 0.10 & 2.35  \\
          & Haute-Provence 5 &            &           &     &  &             \\
          & BH 249           &            &           &     &  &             \\
          & MWSC 2719        &            &           &     &  &             \\
\hline
Terzan 9  & MWSC 2778        & 18 01 38.8 & -26 50 23 &  13.00 $\pm$ 0.10 & 1.76 \\
\hline
Terzan 10 & ESO 521-16       & 18 02 57.4 & -26 04.00 &  13.45 $\pm$ 0.10 & 2.40 \\
          & MWSC 2793        &            &           &     &  &              \\
\hline
Terzan 12 & Terzan 11        & 18 12 15.8 & -22 44 31 &  12.83 $\pm$ 0.10 & 2.06 \\
          & ESO 522-1        &            &           &     &  &             \\
          & MWSC 2838        &            &           &     &  &              \\
\hline
Ton 2     & Pismis 26        & 17 36 10.5 & -38 33 12 &  13.49 $\pm$ 0.05 & 1.24 \\
\hline  
VVVCL002     &                  & 17 41 06.3 & -28 50 42 &  13.80 $\pm$ 0.15 & 2.88 \\
\noalign{\smallskip} 
\hline\end{tabular}
\tablefoot{* From H10 }
\end{table*}

The pipeline provided by ESO (version 2.8) was used to perform the bias, flatfield, distortion correction, the wavelength calibration, extraction and the sky subtraction. IRAF was also used for the combination of the spectra ({\it scombine} task) and the normalization of the combined spectra ({\it continuum} task).

%%%%%%%%%%%%%%%%%%%%%%%%%%%%%%%%%%%%%%
%___________________________________________________________
\section{Radial velocity and equivalent width measurements}
\label{sec:rv}

We measured the radial velocities (RV) of our targets following the method used by our group in previous work employing the CaT \citep[e.g.][]{parisi+15,parisi+16,parisi+22}. The {\it fxcor} task
was used for performing cross-correlation between the observed stars and the spectra of template stars \citep{cole+04}, belonging to Galactic  open and globular clusters.
We adopt as the final RV the average of the cross-correlation results. The correction for the effect introduced by the offset between the star and slit centers
is explained in detail in \citet{parisi+09}.
We obtained a total error of 7.5 km s$^{-1}$ for our RVs, which is the sum in quadrature between the
typical standard deviation of the different cross-correlations (6 km s$^{-1}$) and the error  in centering the image in the spectrograph slit (4.5 km s$^{-1}$).

Equivalent widths (EW) were measured on the normalized combined spectra by fitting a combination of a Gaussian and a Lorentzian function. As shown by several
authors \citep[e.g.][]{cole+04}, such a function reproduces more accurately both the center of the line and the wings. We used the bandpasses from \citet[][hereafter V15]{vasquez+15} (see their Table 1), which
modify the wavelength ranges defined by \citet{armandroff+88}, in order to better fit the wings, and to be fully consistent with the CaT metallicity calibration of V15 and V18 that we will follow here. We have also measured the EW using the original definitions by \citet{armandroff+88} in order to be fully consistent with the metallicity calibration of C04 and DP20 that we will also investigate. EWs were measured with errors estimated to be between $\sim$ 0.1 $-$ 0.5 \AA \space depending on the line and the S/N of the spectra.

%___________________________________________________________
\subsection{Scaling EW to V18}

Even when the same pseudo-continuum and bandpass regions are adopted by two different analyses, small differences may appear between their measured EWs for the same spectrum, as seen in V15, S12 and V18 in comparison with previous work. We
decided to compare our own EW measurements on some spectra from V18 with their EWs before simply applying their calibrations to our EWs to finally derive the metallicities. V18 kindly provided 119 reduced and extracted non-normalised spectra for the clusters Djorg\,2, Terzan\,1, Terzan\,2, Terzan\,8, Terzan\,9, Ton\,2, NGC\,6426, NGC\,6864, and Pal\,10, covering the metallicity range -2.4$\lesssim$[Fe/H]$\lesssim$-0.2.

\begin{figure}[!htb]
    \centering
     \includegraphics[width=\columnwidth]{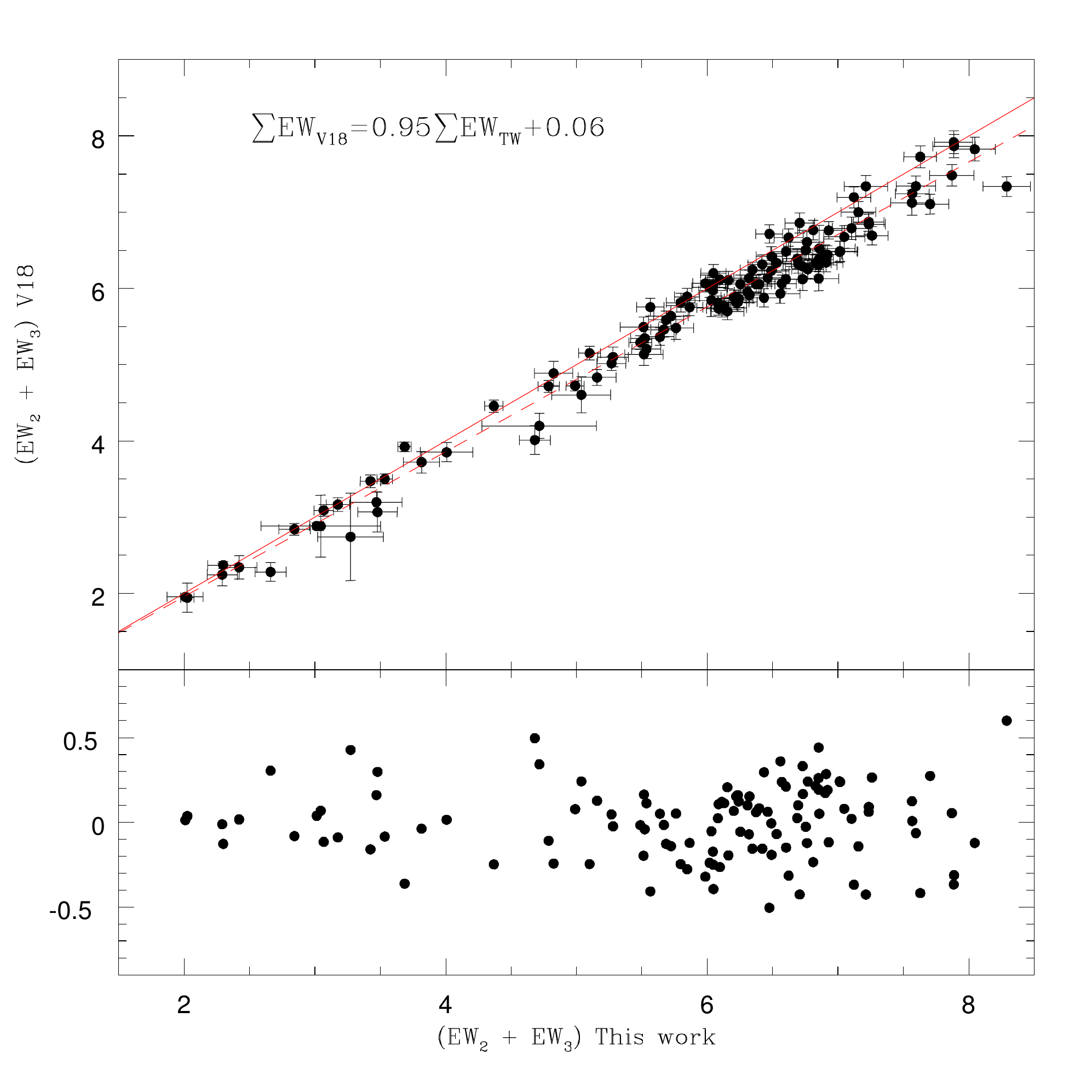}
    \caption{Top: sum of the EW from the two strongest CaT lines as measured by us and by V18 (corrected to the S12 scale). The continuous line shows the one-to-one relation while the dashed line shows the best-fit relation given by Eq. \ref{eq:EW}. 
    Bottom: the residuals of the comparison to the fit. }
    \label{fig:ew}
\end{figure}

We have normalised the spectra of all 119 stars from V18 and measured the EWs with the same pseudo-continuum and bandpass regions and compare our EWs to their's. We find a tight correlation between our measurements (denoted TW for This Work) and theirs (corrected to the S12 scale) when comparing the sum of the two strongest lines, with a rms of 0.2, which translates to a typical metallicity error of 0.12 dex. The relation is given in Eq.\ref{eq:EW} and displayed in Fig. \ref{fig:ew}. 
We then adopt this correction to the sum of the EW of the two strongest CaT lines to calculate metallicities on the scales of V15 and V18. 

\begin{equation}
    \sum{\rm EW_{V18}}=0.95\cdot \sum{\rm EW_{TW}} + 0.06
\label{eq:EW}
\end{equation}

The scaling relation given by Eq. \ref{eq:EW} translates into a small offset in the final metallicities of about 0.03 dex. We only had spectra from V18, therefore the equivalent comparisons for V15 and C04/DP20 were not performed, but based on the above comparison we infer that any additional systematic errors in the final metallicities using the V15 and C04/DP20 scales should be of the order of 0.05 dex or less.

%%%%%%%%%%%%%%%%%%%%%%%%%%%%%%%%%%%%%%
%___________________________________________________________
\section{Metallicity determination and membership}
\label{sec:mets}

It is well known that the sum of the EW of the CaT lines ($\Sigma$EW) correlates with metallicity \citep{armandroff+88}. From this fact, the  CaT
technique was developed and widely used over the last decades. However, since $\Sigma$EW depends not only on metallicity, but also on  effective temperature and surface gravity
\citep{armandroff+91,olszewski+91}, different authors have proposed the use of the so-called reduced EW ($W'$), which removes the dependence on these two last parameters, via its correlation with the magnitude of the observed star or, even better, the difference in magnitude between the observed star and the level of the horizontal branch in a given filter (in our case $K_s-K_{s,HB}$). Many studies have calibrated the $\Sigma$EW with metallicity for different filters (see \citealt{dias+20}, hereafter DP20, for a detailed description of the available calibrations). 
 We decided to use the calibration of V18 that is based on the metallicity scale of \citet{dias+16,dias+16b}, which is the most up-to-date scale for Milky Way globular clusters in the whole metallicity range including metal-rich bulge clusters. Additionally, we also use the calibrations of V15 and DP20 (that follows C04) for comparison purposes.

For our individual targets, we adopted the $K_s$ magnitudes from the VVV survey \citep{Minniti+10}.
For our cluster sample, we followed the procedure described in \citet{mauro+14} in order to calculate the $K_{s,HB}$. 
The magnitude at the HB level was determined by the position of the peak in the luminosity distribution of the reddest part of the HB.
To improve the peak determination, we start from initial guesses obtained from several sources, such as the values presented in \citet{Valenti+07,Valenti+10}. For some clusters, we calculated a ``theoretical'' value based on \citet{bressan+12} and \citet{Girardi+01}, corrected for distance modulus and reddening of the GC.
An empirical value was also calculated from the $V_{HB}$ value listed in H10, corrected for distance modulus and reddening of the GC, and for a mean $(V-K_s)$ color determined using \citet{bressan+12}.
The accuracy of these two methods strongly depends on the accuracy of the known photometric parameters of the cluster.
Unlike the sample of \citet{mauro+14}, the clusters analyzed in this work are affected by higher extinction
and greater differential reddening, and more
contamination by field stars.
To best estimate the peak in the luminosity distribution of the reddest part of the HB, for each cluster we produced color-magnitude (CMD) and Hess diagrams for different cuts in distance from the cluster center (typically 30'', 60'' and 90''). For each cut, we selected comparison fields covering an equal area.
We scrutinized these CMDs and Hess diagrams to determine which overdensities belong to the cluster and which to the surrounding environment, e.g. the Galactic bulge or spiral arms.
For Terzan10, a cluster particularly affected by differential reddening, we based our estimations mainly on the values given by \citet{AlonsoGarcia+15}.

The slope $\beta$ of the relation between magnitude and $\sum$EW,
needed to correct for temperature and luminosity effects, varies with the adopted filter and with the number of lines considered in $\sum$EW. V15 found $\beta_{\rm Ks}=0.384\pm0.019$, which we also use to calculate metallicities on the V15 scale. V18 found $\beta_{\rm V}=0.55$, which is converted to ${\rm K_s}$ using the recipes by DP20, deriving $\beta_{\rm Ks}=0.37$. We note that V18 is on the same scale as S12, and M14 found $\beta_{\rm Ks}=0.385\pm0.013$. Therefore, $\beta$ is very consistent among these calibrations, all using the two strongest CaT lines. In the case of the C04 scale, they use all three CaT lines and the V filter, with $\beta_{\rm V}=0.73\pm0.04$, which was fitted by DP20 resulting in $\beta_{\rm V}=0.71\pm0.05$ and $\beta_{\rm Ks}=0.48\pm0.06$.

 The reduced EW calculated for each star on each scale described above is then converted into metallicity following the respective scales, in order to be fully consistent. V15 derived\\
 ${\rm[Fe/H] = -3.150 + 0.432W'+0.006W'^2}$, V18 derived ${\rm[Fe/H] = -2.68 + 0.13W'+0.055W'^2}$, and \\
 DP20 derived ${\rm[Fe/H] = -2.917 + 0.353W'}$.

In order to discriminate between cluster members and surrounding field stars, we apply the same membership determination method used by our group in previous CaT work \citep{parisi+09,parisi+15,dias+21,parisi+22,dias+22}.
Briefly, we apply criteria including the distance of the star from the center of the cluster, its RV, [Fe/H] and proper motion. To be considered a member, a star must satisfy all of the following: 1) within the adopted cluster radius. We built the radial stellar density profile (see \citealt{parisi+09,parisi+15,parisi+22} for more details) in order to determine the radius;  2) an RV that falls within the error plus intrinsic dispersion (generously adopted as $\pm$ 15 km s$^{-1}$)
from the cluster mean, and ideally different from the average RV of the surrounding stellar field; 3) an [Fe/H] value within the adopted metallicity cuts ($\pm$ 0.20 dex, given by the mean error in the metallicity determinations) of the mean; and 4) a PM that lies within 3 standard deviations of the cluster mean.
We used the proper motions from the Gaia DR3\footnote{\url{https://www.cosmos.esa.int/web/gaia/data-release-3}} survey \citep{gaia+21}. 

\begin{figure}[!htb]
\centering
\includegraphics[width=\columnwidth]{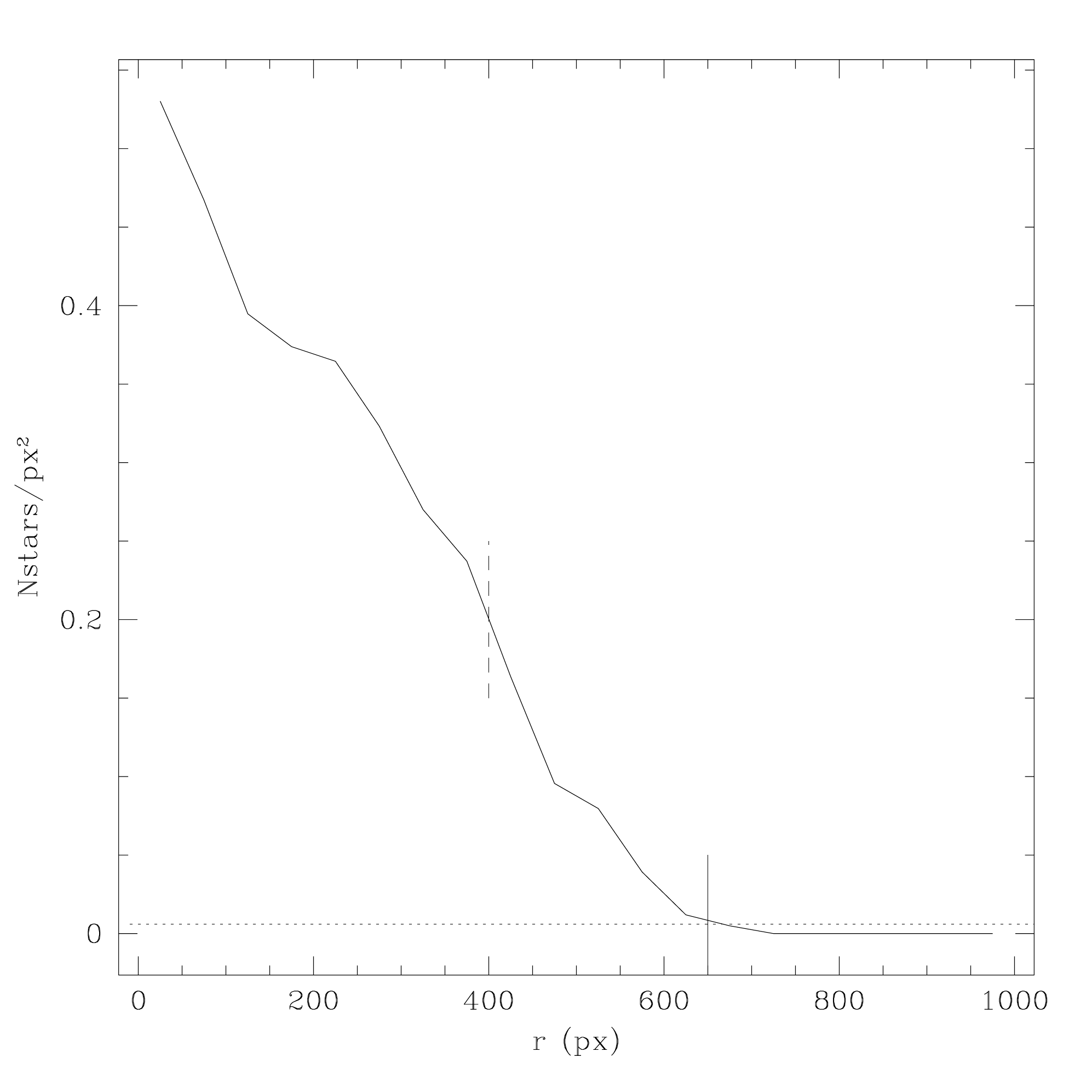}
   \caption{Radial stellar density profile of the cluster NGC\,6401. Horizontal line shows the stellar background level. Solid and dashed vertical lines represent the measured and the adopted cluster radius, respectively. }
      \label{f:profile}
\end{figure}

In Figure \ref{f:N6401_cmd}, we show the CMD of the cluster NGC 6401 as an example, and
in Figure \ref{f:profile} we show the stellar radial density profile of this cluster. By our definition, the radius of the cluster (solid vertical line in the figure) is the point where the stellar density profile intersects the background level (dotted line). For the present analysis, we adopt a more conservative radius (dashed line) to increase the membership probability of the stars adopted as cluster members. We carefully checked that stars with RVs and metallicities compatible with cluster membership had not been discarded due to a very restrictive cutoff in the radius. No additional potential members were found in our sample beyond the adopted radius but within the tidal radius. Note that
the structural parameters for most of these clusters are very poorly known. 
In addition to observational difficulties, many of them are also likely core-collapsed, so they are not well fit by conventional (e.g. \citealt{king66}) analytical profiles. 

\begin{figure}[!htb]
\centering
\includegraphics[width=\columnwidth]{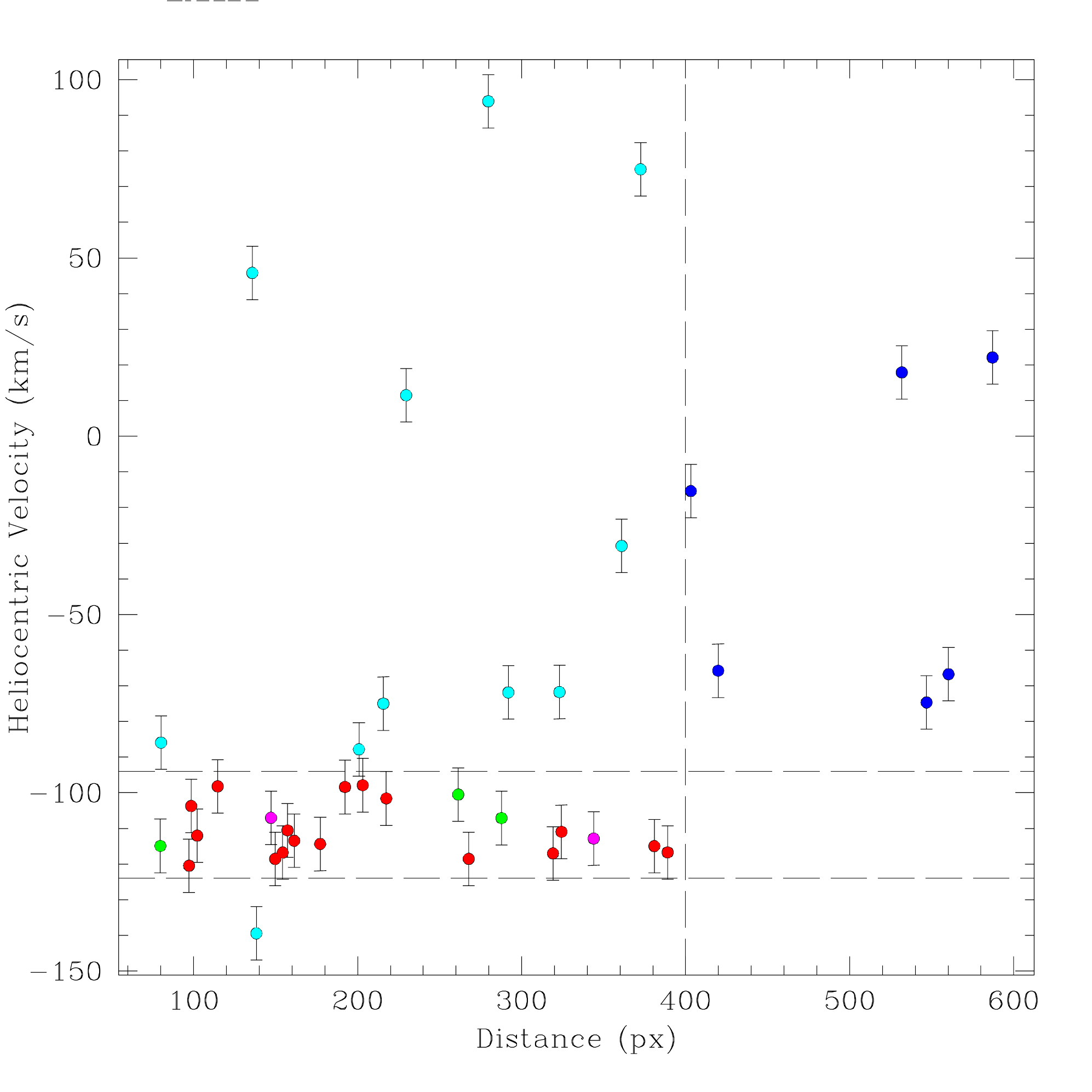}
   \caption{Heliocentric radial velocity vs. distance from the cluster center for NGC\,6401 targets. Symbols as in Figure \ref{f:N6401_cmd}. Radial velocity error cuts ($\pm$ 15 km s$^{-1}$, horizontal lines) and the adopted cluster radius (vertical line) are shown.
   }
      \label{f:rv_r}
\end{figure}

\begin{figure}[!htb]
\centering
\includegraphics[width=\columnwidth]{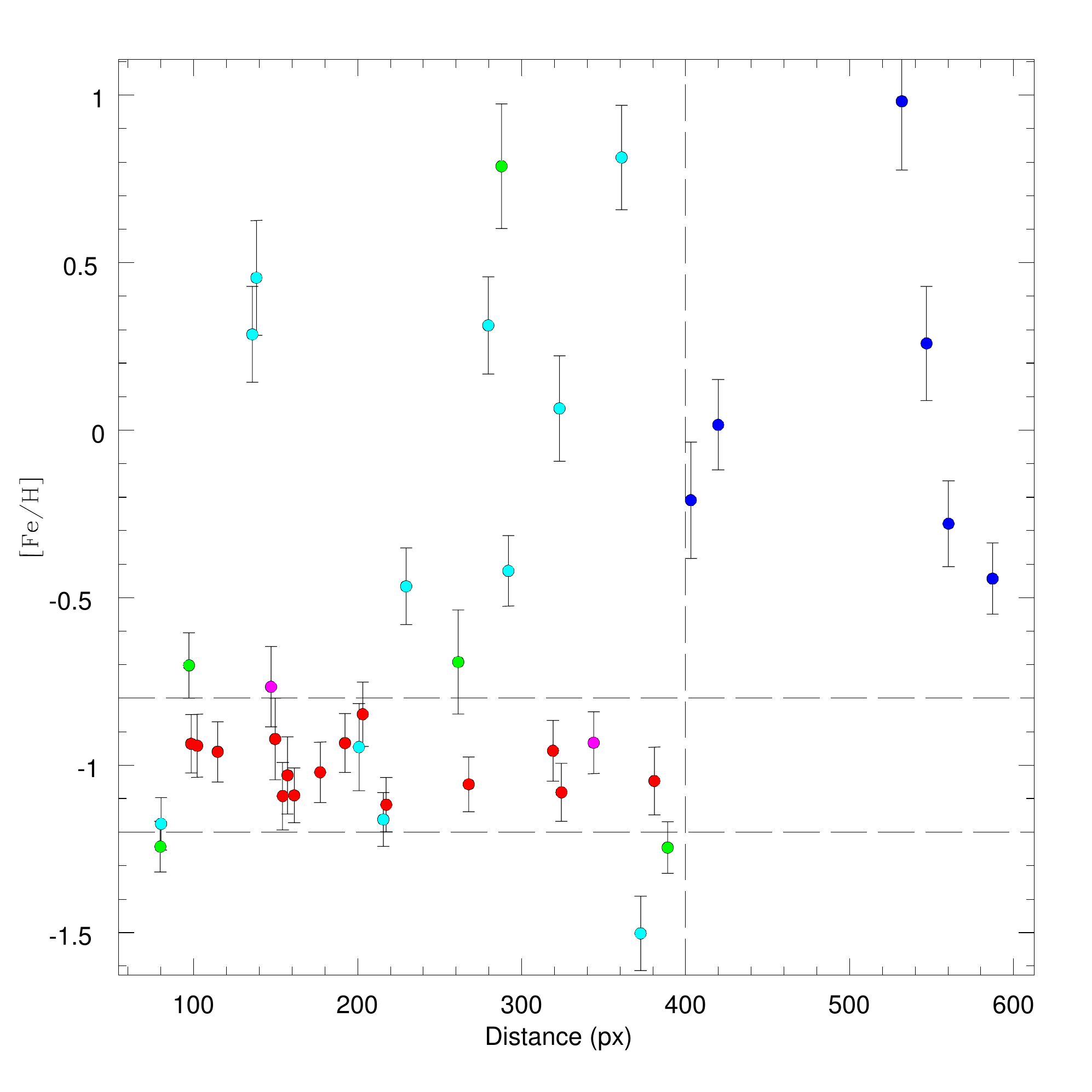}
   \caption{ Metallicity  vs. distance from the cluster center for NGC\,6401 targets. The color code is the same as in figure \ref{f:N6401_cmd}. Metallicity error cuts ($\pm$ 0.2 dex, horizontal lines) and the adopted cluster radius (vertical line) are shown. 
   }
      \label{f:met_r}
\end{figure}

\begin{figure}[!htb]
\centering
\includegraphics[width=\columnwidth]{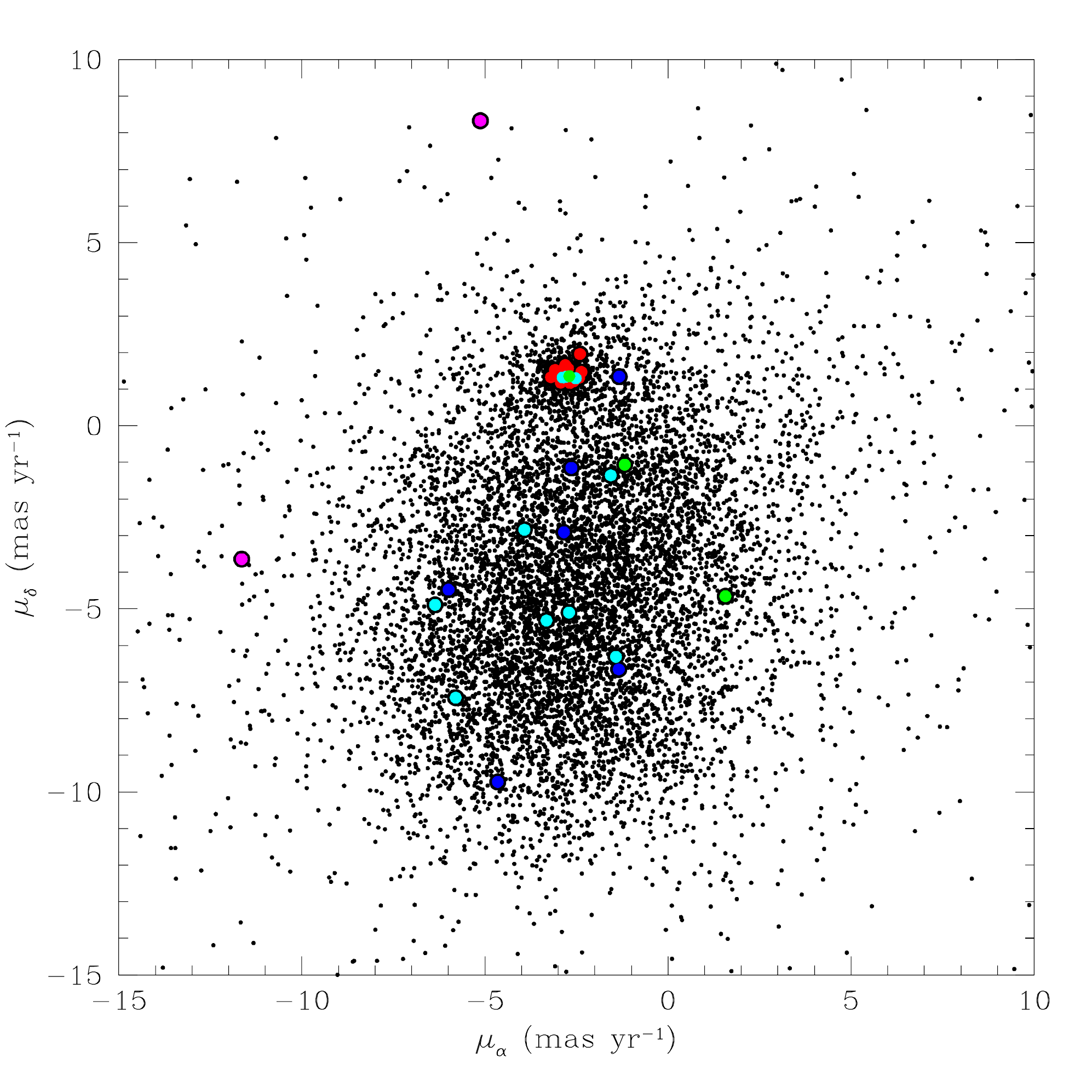}
   \caption{Proper motion plane for the cluster NGC\,6401. Black points represent stars from the Gaia eDR3 catalogue and large circles stand for  our spectroscopic targets. The color code is the same as in Figure \ref{f:N6401_cmd}.
   }
      \label{f:pm}
\end{figure}

For the same  example cluster, we include in Figures \ref{f:rv_r} and \ref{f:met_r} the behavior of RV and metallicity with distance from the center, respectively. Figure \ref{f:pm} show the positions of our targets in the PM plane from Gaia eDR3. The color code used in these figures is the same as  in Fig.\ref{f:N6401_cmd} and in our previous CaT work (see for example \citealt{parisi+22,dias+22}): field stars located at a distance from the cluster center larger than the adopted radius have been plotted with blue symbols; stars discarded because they have RVs or  metallicities outside of the adopted cuts are shown in cyan and green, respectively; magenta symbols represent stars discarded because of their discrepant PMs and red circles represent stars that have passed all criteria and, therefore, are considered our final cluster members. 

Unfortunately, the VVV CL002 observations were made long before Gaia, and none of our sample of 14 stars observed in this area passed our membership criterion. In particular, all stars fall well away from the
cluster mean PM found by Vasiliev and Baumgardt (2021). This is a graphic illustration of the importance of PM in determining cluster membership for such crowded and convolved fields. Thus, VVV CL002 is excluded from further discussion. However, this remains a very interesting target, as \citet{minniti+21} confirmed that this is a real GC based on the VVV PM diagram, and concluded that it is the closest GC to the Galactic center.

We found a total of 130 members in  our remaining 12 clusters, for an average of 11 members per cluster, with the range falling from only 2 - 3 stars (in BH 261, Djorg 2 and Terzan 6) to 19 in NGC 6642. 
For our members, we include in Table \ref{t:individuals}, consecutively, the star identification, the equatorial coordinates, RV, $K_s-K_s(HB)$, $\Sigma$EW (for the two strongest lines) and metallicity (corresponding to the V18 calibration), with their respective errors. 

\begin{table*}
\caption{Measured Values for Member Stars}             % title of Table
\label{t:individuals}      % is used to refer this table in the text
\centering                          % used for centering table
\footnotesize
\begin{tabular}{lccccccc}
\hline
\multicolumn{1}{c}{Star $^*$} & 
\multicolumn{1}{c}{R.A.} & 
\multicolumn{1}{c}{Dec} & 
\multicolumn{1}{c}{RV} & 
\multicolumn{1}{c}{$K_s-K_s(HB)$} & 
\multicolumn{1}{c}{$\Sigma$EW} & 
\multicolumn{1}{c}{[Fe/H]} \\
\multicolumn{1}{c}{} &
\multicolumn{1}{c}{J2000} & 
\multicolumn{1}{c}{J2000}  & 
\multicolumn{1}{c}{km s$^{-1}$}  & 
\multicolumn{1}{c}{mag} & 
\multicolumn{1}{c}{\AA} & 
\multicolumn{1}{c}{dex} \\
\hline
\noalign{\smallskip}
BH261$-$286607$-$M$-$09 & 18.2345  & -28.6465  & -52.5 $\pm$ 1.9  & -0.38 $\pm$ 0.10  & 4.34 $\pm$ 0.19 & -1.25 $\pm$ 0.11  \\
BH261$-$291014$-$M$-$16 & 18.2352  & -28.6349  & -42.23 $\pm$ 1.2 & -2.23 $\pm$ 0.10  & 5.02 $\pm$ 0.09 & -1.27 $\pm$ 0.08 \\
BH261$-$320470$-$M$-$24 & 18.2358  & -28.6232  & -39.9 $\pm$ 2.3  & -0.62 $\pm$ 0.10  & 4.74 $\pm$ 0.16 & -1.08 $\pm$ 0.11  \\      
\noalign{\smallskip} 
\hline\end{tabular}
\tablefoot{ This Table is available in its entirety in the online journal and in the CDS database. A portion is shown here for guidance regarding its form and content. $^*$ Cluster name - ID from the photometry - CCD chip (M:master, S: secondary) -  aperture number.}
\end{table*}

\begin{figure}[!htb]
\centering
\includegraphics[width=\columnwidth]{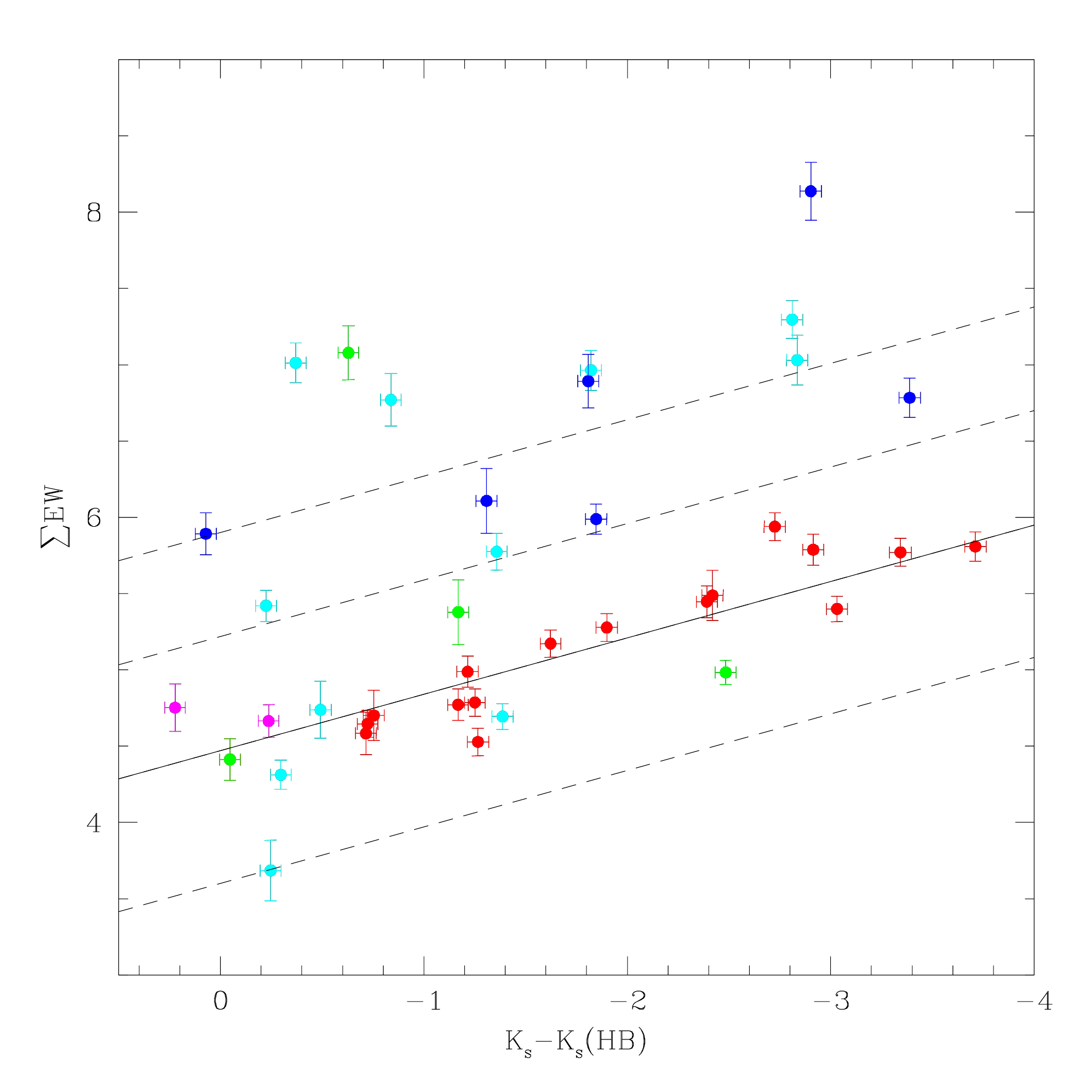}
   \caption{The sum of the equivalent width of the three CaT lines vs. the difference  $K-K_{HB}$ for stars identified as members of NGC\,6401. The color code is the same as in Figure \ref{f:N6401_cmd}. The solid line represents a metallicity of -1.0, while dashed lines represent [Fe/H] = 0.0,-0.5, and -1.5, from top to bottom. }
  \label{f:slope_6401}
\end{figure}

\begin{figure}[!htb]
\centering
\includegraphics[width=\columnwidth]{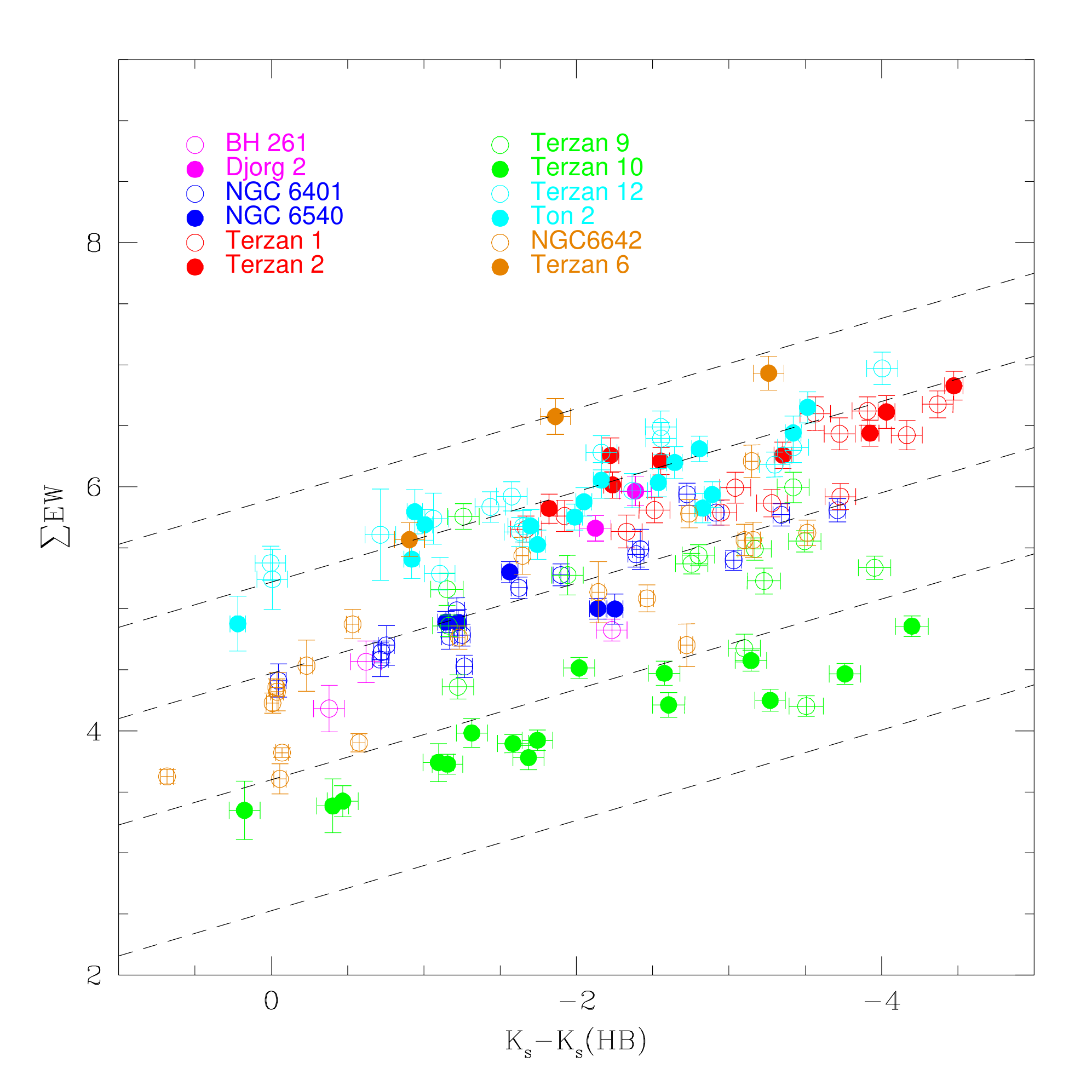}
   \caption{The same as in Figure \ref{f:slope_6401} but for members in all clusters. Dashed lines represent [Fe/H] = 0.0, -0.5, -1.0, -1.5 and -2.0 from top to bottom.  
   }
      \label{f:slopes}
\end{figure}

In Figure \ref{f:slope_6401} we show
the behavior of $\Sigma$EW as a function of $K-K_{s,HB}$ for targets in NGC 6401, where one sees that cluster members follow an iso-abundance line, with a mean metallicity of  -1.0. %-0.91. 
The same can be appreciated in Figure \ref{f:slopes} for all clusters in our sample, considering only cluster members for each cluster individually. For each cluster, red giants follow lines of equal slope but different zero points, which depend only on the cluster metallicity. This graphically displays the power of the CaT technique to derive metallicity.

Finally, we  calculated the cluster mean RV and metallicity on  the adopted V18 scale, and on two additional scales - V15 and DP20 - for comparison. The results along with 
their corresponding errors  
are presented in Table \ref{t:results}.
We also include the number of members in each cluster and the mean PM derived, which are in very good agreement with the mean values derived by \citet{vasiliev+21} considering the errors.  The mean metallicity is typically determined to an internal error of 0.05 dex, while the mean RV has a mean error of 3km/s.
None of our clusters show strong evidence for a range in metallicities significantly exceeding that expected from measurement errors, although NGC 6642 and Terzan 6 have a substantially larger range than the other clusters. For the rest of the paper, we will utilize our mean metallicity on the  V18 scale, following the \citet{dias+16,dias+16b} metallicity
scale as our favored value, as delineated above.
 
\begin{table*}
\caption{Derived cluster mean  parameters }         % title of Table
\label{t:results}      % is used to refer this table in the text
\centering                          % used for centering table
\footnotesize
\begin{tabular}{lrrrrrrr}
\hline
\multicolumn{1}{c}{} &
\multicolumn{1}{c}{n} & 
\multicolumn{1}{c}{RV} & 
\multicolumn{1}{c}{[Fe/H]$_{\rm V18}^{\rm (adopted)}$} &
\multicolumn{1}{c}{[Fe/H]$_{\rm V15}$} &
\multicolumn{1}{c}{[Fe/H]$_{\rm DP20}$} &
\multicolumn{1}{c}{$\mu_{\alpha}$} &
\multicolumn{1}{c}{$\mu_{\delta}$}\\  
\multicolumn{1}{c}{} & 
\multicolumn{1}{c}{} & 
\multicolumn{1}{c}{km s$^{-1}$} &
\multicolumn{1}{c}{dex} &
\multicolumn{1}{c}{dex} & 
\multicolumn{1}{c}{dex} &
\multicolumn{1}{c}{mas yr$^{-1}$} &
\multicolumn{1}{c}{mas yr$^{-1}$}\\
\hline
BH\,261        & 3    &  -44.9 $\pm$ 3.8(6.7)  &  -1.21 $\pm$ 0.06(0.10)    & -1.19 $\pm$ 0.05(0.09)    & -1.09 $\pm$ 0.05(0.09)   & 3.52  $\pm$ 0.11(0.18) & -3.63 $\pm$ 0.16(0.27) \\
Djorg\,2       & 2    & -162.4 $\pm$ 9.1(12.9) &  -0.67 $\pm$ 0.07(0.10)    & -0.75 $\pm$ 0.05(0.08)    & -0.70 $\pm$ 0.05(0.07)   & 0.65  $\pm$ 0.03(0.04) & -3.00 $\pm$ 0.02(0.03) \\
NGC\,6401      & 15   & -110.8 $\pm$ 1.8(7.8)  &  -1.00 $\pm$ 0.03(0.12)    & -1.01 $\pm$ 0.02(0.10)    & -0.91 $\pm$ 0.02(0.09)   & -2.75 $\pm$ 0.06(0.24) &  1.43 $\pm$ 0.05(0.20) \\
NGC\,6540      & 5    &  -22.1 $\pm$ 1.3(2.9)  &  -1.04 $\pm$ 0.06(0.14)    & -1.05 $\pm$ 0.05(0.11)    & -0.98 $\pm$ 0.04(0.10)   & -3.74 $\pm$ 0.04(0.08) & -2.74 $\pm$ 0.10(0.21) \\
NGC\,6642      & 19   &  -48.2 $\pm$ 1.8(8.0)  &  -1.11 $\pm$ 0.06(0.24)    & -1.11 $\pm$ 0.05(0.21)    & -1.04 $\pm$ 0.04(0.15)   & -0.21 $\pm$ 0.04(0.18) & -3.86 $\pm$ 0.05(0.20) \\
Terzan\,1      & 13   &   68.4 $\pm$ 3.1(11.2) &  -0.71 $\pm$ 0.04(0.14)    & -0.77 $\pm$ 0.03(0.11)    & -0.73 $\pm$ 0.03(0.11)   & -2.84 $\pm$ 0.08(0.28) & -4.87 $\pm$ 0.10(0.37) \\
Terzan\,2      & 8    &  130.5 $\pm$ 2.3(6.5)  &  -0.54 $\pm$ 0.03(0.10)    & -0.65 $\pm$ 0.03(0.07)    & -0.56 $\pm$ 0.02(0.06)   & -2.14 $\pm$ 0.03(0.09) & -6.31 $\pm$ 0.05(0.13) \\
Terzan\,6      & 3    &  134.7 $\pm$ 3.4(6.0)  &  -0.21 $\pm$ 0.15(0.25)    & -0.42 $\pm$ 0.10(0.18)    & -0.43 $\pm$ 0.08(0.15)   & -4.83 $\pm$ 0.14(0.25) & -7.25 $\pm$ 0.09(0.16) \\    
Terzan\,9      & 14   &   70.1 $\pm$ 2.4(9.0)  &  -1.15 $\pm$ 0.03(0.12)    & -1.14 $\pm$ 0.03(0.11)    & -1.06 $\pm$ 0.03(0.10)   & -2.16 $\pm$ 0.07(0.25) & -7.62 $\pm$ 0.06(0.21) \\
Terzan\,10     & 16   &  209.3 $\pm$ 3.1(12.5) &  -1.64 $\pm$ 0.02(0.09)    & -1.59 $\pm$ 0.02(0.08)    & -1.47 $\pm$ 0.02(0.08)   & -6.99 $\pm$ 0.12(0.47) & -2.56 $\pm$ 0.07(0.28) \\
Terzan\,12     & 16   &  107.4 $\pm$ 1.7(6.7)  &  -0.48 $\pm$ 0.04(0.15)    & -0.60 $\pm$ 0.03(0.11)    & -0.58 $\pm$ 0.03(0.10)   & -6.42 $\pm$ 0.05(0.21) & -3.09 $\pm$ 0.04(0.15) \\
Ton\,2         & 16   & -180.8 $\pm$ 2.0(8.1)  &  -0.57 $\pm$ 0.03(0.13)    & -0.67 $\pm$ 0.02(0.10)    & -0.61 $\pm$ 0.03(0.10)   & -5.97 $\pm$ 0.04(0.15) & -0.81 $\pm$ 0.03(0.12) \\
\noalign{\smallskip} 
\hline\end{tabular}
\tablefoot{Errors correspond to the standard error of the mean (values in parentheses are the standard deviation.) }
\end{table*}

%%%%%%%%%%%%%%%%%%%%%%%%%%%%%%%%%%%%%%
%___________________________________________________________
\section{Comparison with previous results}
\label{sec:lit}

All of our clusters have had previous determinations of both RV and metallicity. Radial velocities 
only require, at a minimum, rather low resolution, low S/N spectra, while metallicities can be derived from a large number of techniques, ranging from broad band photometry to high resolution, high S/N spectroscopy, and integrated light to individual stars, with a concomitantly wide range of accuracy. We therefore expect that RVs for our sample should generally be in reasonable agreement, although of course the possibility of previous studies including non-members is an issue.
However, although published metallicities exist for all of our sample, they indeed come from a variety of methods, many of which are of rather low quality, making these values both inhomogeneous and quite uncertain in general, especially given the fact that our sample has been left relatively unstudied for good reason - namely the high extinction as well as crowding associated with BGCs. We therefore anticipate more significant metallicity discrepancies in our sample.

Here we compare our results with previous literature values for each cluster in turn. We concentrate on RV and [Fe/H] determinations from five (generally) internally homogeneous, recent and widely-used  catalogues, which should help minimize errors associated with field star inclusion and/or lower quality techniques. These include H10 (which in fact is not internally homogeneous or recent but is regarded as the bible of Galactic GC properties),  \citet[][V18]{vasquez+18}, who used the same CaT technique on a large sample of reddened GCs, \citet[][B19]{Baumgardt+19}, a catalog of various GC parameters including RV, Dias (2019, D19)\footnote{https://www.sc.eso.org/~bdias/catalogues.html}, who 
 averaged spectroscopic metallicities from multiple studies when available, after bringing them to a homogeneous metallicity scale - that of
\citet{dias+16,dias+16b},
and Geisler et al. (2022, in preparation - G22), who are studying BGCs with the APOGEE spectrograph in SDSS-IV via the CAPOS (bulge Cluster APOgee Survey) project. We remark that our sample generally included almost twice as many stars per cluster as either V18 or G22, with the additional benefit that V18 did not have Gaia PMs to help select members. 
We note that CAPOS is the only source of high resolution spectroscopic metal abundances (and RVs) for our sample except for Terzan 1, as noted below, while the metallicities from H10 generally come from photometry or, at best, low resolution spectroscopy. We also note that we use the D16 calibration values from V18.  There are no clusters in common with the MUSE CaT sample by \citet{husser+20} for direct comparison, but they follow \citet{dias+16,dias+16b} therefore their results should be on the same scale as ours. All values are listed 
in Table \ref{tab:lit}.
V18 and G22 also include references to other, generally older and less reliable, RV and metallicity derivations for clusters in common with our sample, which are not discussed in detail here.

{\bf BH 261}. Our sample is small, with only 3 members. The only published RV measurement, from B19, is in reasonable agreement with our value. All three [Fe/H] values are very similar. No previous CaT or high resolution studies exist.

{\bf Djorg 2}. Good to reasonable agreement exists with all previous RV values. Our metallicity is in excellent agreement with H10, but much higher than other determinations, with an offset of about 0.4 to 0.45 dex with respect to the V18 CaT and G22 CAPOS values, respectively. However, note that we measured only 2 members, the smallest number in our sample, while V18 had 3 and G22 6 members.

{\bf NGC 6401}. The H10 RV is much higher than our value, which is instead very consistent with that of V18 and B19. Our metallicity value is in good agreement with that of H10 and D19 and in reasonable accord with V18.

{\bf NGC 6540}. All RV values are in good accord, as are [Fe/H] values except that of H10, which is in relatively poor agreement. However, this value is based only on the slope of the RGB in the near IR, a technique of low precision.

{\bf NGC 6642}. RVs are in good accord, although the B19 value falls 15km/s higher than ours. Metal abundances are all in good to excellent agreement.

{\bf Terzan 1}. 
This is a very interesting GC because it is one of the few second-parameter GCs within the bulge, with a red horizontal branch but a steep red giant branch indicating a rather low metallicity. It has been the subject of several recent studies, which makes it a good target for comparison amongst different techniques, both in  RV as well as metallicity.

The first RV published for Terzan\,1 was $35\ {\rm km\ s^{-1}}$, based on integrated CaT spectroscopy \citep{armandroff+88}. 
We suspect this relatively low value compared to all subsequent measurements is due to field star contamination.
The \cite{valenti+15} high-resolution spectroscopy gave an average RV = $57\pm1.8\ {\rm km\ s^{-1}}$, which is offset by about $11\ {\rm km\ s^{-1}}$ from our value. \cite{vasquez+18} derived $63\pm0.5(8)\ {\rm km\ s^{-1}}$, based on 9 stars, which is consistent with both the high-resolution result and our value.

\citet{baumgardt+18} analysed our FORS2 data for Terzan\,1. 

They reduced the spectra, measured RVs and estimated cluster membership based only on RV information, deriving a mean of RV = $57.7\pm1.2\ {\rm km\ s^{-1}}$, more than 10 km/s lower than our value of RV = $68.4\pm3.1\ {\rm km\ s^{-1}}$, which is surprising given the identical data and estimated errors.
We find that there are a number of stars with similar RV as the confirmed members but with different metallicities and proper motions. Therefore, \citet{baumgardt+18} ended up including a number of non-members, since they did not take into account  metallicity and proper motion, which we used in our membership selection. This incompatible membership selection likely  explains the difference in the mean cluster RV obtained by the two studies based on the same data (see Tables \ref{t:results} and \ref{tab:lit}).  
This comparison shows the danger of defining GC membership using only RV. In cases where the contrast with the field is small, the average RV will not suffer much
but of course metallicity could, and in the case of low-mass GCs in a dense field, RV alone is not enough for assessing membership. Finally, \cite{idiart+02} derived a mean RV = $114\pm14\ {\rm km\ s^{-1}}$, which is what the H10 value is based on. This has a very significant offset with respect to all other values. \cite{valenti+15} had member stars in common with \cite{idiart+02} with similar metallicities but discrepant RV, and argued that the RV offset is probably related to some systematic errors not accounted for in \cite{idiart+02}. 

\cite{ortolani+99} and \cite{Valenti+10} derived a metallicity around -1.1 from optical and near ir photometry, respectively.
The metallicity given by H10 of [Fe/H] = -1.03$\pm0.03$ dex agrees with these values. It comes from the average of the integrated low-resolution CaT metallicity by \cite{armandroff+88} of [Fe/H] = -0.71$\pm0.15$ dex on the \cite{zinn+84}  metallicity scale or [Fe/H] = -0.68$\pm0.15$ dex in the \cite{carretta+09} metallicity scale, and the average low-resolution optical spectroscopic metallicity from 7 RGB stars of -1.27$\pm0.05$ dex by \cite{idiart+02}. More recently,  \cite{valenti+15} used high-resolution H-band spectroscopy to find  -1.26$\pm0.03$ dex based on 15 members, some of them in common and in agreement with \cite{idiart+02}. \cite{vasquez+18} used the same technique as we do here and presented final average metallicities based on 9 stars:
[Fe/H]$_{\rm D16}$ = -0.74$\pm0.18$ dex,
which resembles the higher metallicities derived from integrated CaT spectra, as we also find here: $\mathrm{{\rm[Fe/H]}_{\rm V18} = -0.71\pm0.04}$ dex. \cite{vasquez+18} argued that the HB magnitude was very uncertain in the V filter because of differential reddening, increasing the uncertainty in metallicity to about 0.15 dex. In summary, three studies based on CaT agree on a higher metallicity than that found by photometry and high-resolution spectroscopy. Potentially, the difference could be related to the high $\alpha$-element abundance of this GC ($\sim0.4$ dex, \citealp{valenti+15}), that makes the overall metallicity increase by about $\sim$0.2 dex with respect to [Fe/H], i.e., [M/H] = -0.91 \citep{Valenti+10}.

{\bf Terzan 2}. Our RV is in excellent agreement with B19 and G22, rough agreement with V18 and poor agreement with H10. Note that V18 only included 3 members while we have 8. We find good accord with other metallicities except for G22, which is again about 0.35 dex lower than our determination.

{\bf Terzan 6}. Our sample is small, with only 3 members. Very good agreement exists with both previous RV determinations. However, our [Fe/H] value is more than 0.3 dex higher than the H10 or D19 values. Both of these metallicities are based only on the near IR RGB slope.  The largest metallicity difference among the three scales in Table \ref{t:results} is for this cluster, amounting to 0.22dex, and this is also the highest metallicity cluster in our sample by a substantial amount. The difference is expected because the calibration by V18, based fully on globular clusters by \citet{dias+16,dias+16b}, differs from V15 only in the metal-rich regime. The argument in V18 was that their calibration is based only on globular clusters including metal-rich bulge GCs, whereas V15 relied on bulge field stars for the metal-rich regime. Therefore we keep our choice for the metallicity following V18 in Table \ref{t:results}.

{\bf Terzan 9}. Very good agreement is found with all RV values, with the notable exception of B19, whose value is more than 40 km s$^{-1}$ lower than ours. Excellent 
metallicity agreement exists, with the notable exception of G22, which again is about 0.25 dex lower than our determination. We note that \citet{ernandes+19} have carried out low resolution VLT-MUSE observations and derived a mean RV of $58.1\pm 1.1 {\rm km\ s^{-1}}$ and mean [Fe/H] = -1.10$\pm 0.15$ from a large number of stars, in good agreement with our values. 

{\bf Terzan 10}. The only previous RV measurement is that of B19, whose value of -64.11$\pm$3.09 \rm km s$^{-1}$ is over 270 km/s lower than ours. This is by far the worst discrepancy in our comparison. We note that our sample includes 16 members and that our error is very reasonable, and also that we do not find any stars with RVs between 0 and -100km/s in our sample of 46 stars (including the secondary chip). Thus, we strongly suspect that there is an error in the B19 value. 
 In fact, they have recently updated their determination in their personal 
website\footnote{\url  https://people.smp.uq.edu.au/HolgerBaumgardt/globular/fits/ter10.htm} to $211.37\pm2.26\ {\rm km\ s^{-1}}$, which is now in very good agreement with our determination.
Metallicity agreement is also the worst of all our clusters, with a discrepancy of almost 0.7 dex. However, both the H10 and D19 values are based only on the RGB slope in the V-I color, which is even more susceptible to errors such as reddening than the near IR technique. 
This cluster suffers from extreme differential extinction, as seen in \citet{AlonsoGarcia+15} and \citet{cohen+18}, which is undoubtedly at least partly to blame for the above discrepancies.
 Moreover, \citet{AlonsoGarcia+15} studied variable stars in Terzan\,10 and found that based on their periods combined with the [Fe/H]$\sim -1.0$ from H10, this cluster lies between Oosterhoff groups II and III. Spectroscopic metallicity was required for a definitive classification: [Fe/H]$\sim -0.5$ would mean a rare case of Oosterhoff III, and [Fe/H]$\lesssim -1.5$ would mean Oosterhoff II. We find [Fe/H]$= -1.64$, finally resolving this uncertainty, as already mentioned by \citet{alonso-garcia+21} using our preliminary results.

{\bf Terzan 12}. Reasonable RV agreement is found, as well as excellent metallicity agreement, with the H10 and D19 values, based only on the VI RGB slope.

{\bf Ton 2}. We find very good accord amongst the variety of RV derivations, while the metallicity values are either close to ours (H10 and G22) or about 0.3 dex higher. We note that  \citet{fernandez-trincado+22} find very similar results to G22 from essentially the same sample. 

Comparing our RVs to previous determinations, we find a mean difference (in the sense our value - previous) of -3.2$\pm$ 25.1 km s$^{-1}$ for 9 clusters in common with H10, 
-2.8$\pm 7.5 km s^{-1}$ for 6 clusters in common with V18,
$+23.3\pm 80.3 km s^{-1}$ for 12 clusters in common with B19,
and -2.9$\pm 6.2 km s^{-1}$ for 6 clusters in common with G22. All of these differences are quite reasonable except with B19, which in fact is now generally regarded as the most reliable compilation. However, as noted above, there are 2 very strong outliers in our comparison with B19: Terzan 9 and 10. If we eliminate these two, we find a mean difference of only -3.4$\pm 10.6 km s^{-1}$ for the other 10 clusters. Thus, overall agreement with prior published values is good, but does suggest our values are about 3 km/s too low on average. Another explanation is that the uncertainties are slightly underestimated.  

As for metallicities, we find a mean difference (in the same sense as above, adopting [Fe/H]$_{\rm V18}$) of $+0.07\pm$ 0.25 dex for 12 clusters in common with H10, 
0.00$\pm$ 0.20 dex for 6 clusters in common with V18,
$-0.06\pm$ 0.25 dex for 12 clusters in common with D19,
and $+0.22\pm 0.16$dex for 6 clusters in common with G22.
 We stress the very good agreement with V18, 
 as expected, because by construction our metallicity determinations are on the same scale, which is generally compatible with high-resolution spectroscopy and with H10. This is also true of D19 but the agreement is not as good.
The large 
offset from G22  is somewhat surprising. 
We emphasize again that our sample size is on average almost twice that of 
G22 for the clusters in common.
However, G22 did  have the benefit of Gaia proper motions as a membership criterion, and also use high resolution spectroscopy, which should yield more robust metallicities compared to our CaT technique. 
 We notice however that the large offset is the average of a small offset of +0.11 $\pm$ 0.12 dex from 3 clusters and a large offset of +0.32 $\pm$ 0.13 dex from the other 3 clusters, therefore, the large offset may be due to something particular to these 3 clusters.
 Indeed, 2 of them are by far the most reddened of the six.
\citet{nidever+20} compare the metallicities
of stars in 26 GCs with APOGEE ASPCAP metallicities ranging from -0.6 to -2.3 with those of other high-resolution studies, and
found a mean offset of 0.06 dex to higher metallicity for APOGEE
and a scatter of 0.09 dex, while \citet{fernandez-trincado+20}
find an offset of 0.11 $\pm 0.11$dex in the opposite sense when comparing ASPCAP to BACCHUS abundances. Thus, the cause of this discrepancy  may be related to different techniques to analyse APOGEE spectra combined with some challenges in three particular GCs. 

\begin{table}[]
    \centering
    \footnotesize
        \caption{RV and metallicity from five internally homogeneous GC catalogues for our BGCs}
    \label{tab:lit}
    \begin{tabular}{c|ccc}
\noalign{\smallskip}
\hline
\noalign{\smallskip}
Name  & RV (${\rm km\ s^{-1}}$) & [Fe/H] & Source  \\
\noalign{\smallskip}
\hline \hline
\noalign{\smallskip}
BH\,261 & --- & -1.30        & H10  \\
“         & -29.38$\pm0.60$ & ---            & B19  \\
“       & --- & -1.27$\pm0.16$ & D19** \\
\noalign{\smallskip}
Djorg\,2 & --- & -0.65        & H10  \\
“       & -159.9$\pm0.9$ & -0.97$\pm0.13$            & V18 \\
“         & -148.05$\pm1.38$ & ---            & B19 \\
“        & --- & -0.91$\pm0.05$ & D19  \\
“       & -152.0$\pm1.2$ & -1.14$\pm0.04$            & G22  \\
\noalign{\smallskip}
NGC\,6401 & -65.0$\pm8.6$ & -1.02        & H10 \\
“       & -115.4$\pm0.8$ & -1.18$\pm0.14$            & V18  \\
“         & -99.26$\pm3.18$ & ---            & B19  \\
“         & ---           & -1.08$\pm0.06$ & D19  \\
\noalign{\smallskip}
NGC\,6540 & -17.7$\pm1.4$ & -1.35        & H10  \\
“         & -17.98$\pm0.84$ & ---            & B19  \\
“         & ---           & -0.89$\pm0.73$ & D19  \\
“       & -14.4$\pm1.1$ & -1.09$\pm0.06$            & G22  \\
\noalign{\smallskip}
NGC\,6642 & -57.2$\pm5.4$ & -1.26        & H10  \\
“         & -33.23$\pm1.13$ & ---            & B19  \\
“         & ---           & -1.03$\pm0.17$ & D19  \\
“       & -55.4$\pm2.4$ & -1.11$\pm0.04$            & G22  \\
\noalign{\smallskip}
Terzan\,1 & $114\pm14$   & -1.03        & H10  \\
“       & $63.0\pm1.5$ & -0.74$\pm0.18$            & V18  \\
“         & $57.55\pm1.61$ & ---            & B19  \\
“         & ---          & -0.74$\pm0.09$ & D19  \\
\noalign{\smallskip}
Terzan\,2 & $109.0\pm15.0$ & -0.69        & H10  \\
“       & $144.6\pm1.4$ & -0.42$\pm0.18$            & V18  \\
“         & $128.96\pm1.18$ & ---            & B19  \\
“         & ---        & -0.42$\pm0.21$ & D19  \\
“       & $134.1\pm1.1$ & -0.88$\pm0.02$            & G22  \\
\noalign{\smallskip}
Terzan\,6 & $126.0\pm15.0$    & -0.56        & H10  \\
“         & $137.15\pm1.7$ & ---            & B19  \\
“         & ---           & -0.53$\pm0.16$ & D19**  \\
\noalign{\smallskip}
Terzan\,9 & $59.0\pm10.0$ & -1.05        & H10  \\
“       & $71.4\pm1.0$ & -1.08$\pm0.14$            & V18  \\
“         & $29.31\pm2.96$ & ---            & B19  \\
“         & ---        & -1.08$\pm0.16$ & D19  \\
“       & $69.8\pm5.1$ & -1.42$\pm0.04$            & G22  \\
\noalign{\smallskip}
Terzan\,10 & ---       & -1.00        & H10  \\
“         & -64.11$\pm3.09$ & ---            & B19  \\
“          & ---       & -0.97$\pm0.16$ & D19**  \\
\noalign{\smallskip}
Terzan\,12 & $94.1\pm1.5$ & -0.50        & H10  \\
“         & $94.77\pm0.97$ & ---            & B19  \\
“          & ---          & -0.47$\pm0.16$ & D19**  \\
\noalign{\smallskip}
Ton\,2 & -184.4$\pm2.2$  & -0.70        & H10  \\
“       & -172.7$\pm0.8$ & -0.26$\pm0.15$            & V18  \\
“      & -184.72$\pm1.12$ & ---            & B19  \\
“      & ---            & -0.26$\pm0.27$ & D19  \\
“       & -177.9$\pm4.0$ & -0.73$\pm0.03$            & G22  \\
\noalign{\smallskip}
\hline
\noalign{\smallskip}
    \end{tabular}\\
     Notes: H10: \citet{harris10}, V18: \citet[][in D16 scale]{vasquez+18}, B19: \citet{Baumgardt+19}, D19: Dias compilation, and G22: Geisler et al. in prep. **taken from H10 with an offset, i.e., it is not a spectroscopic metallicity.
\end{table}

%%%%%%%%%%%%%%%%%%%%%%%%%%%%%%%%%%%%%%
%___________________________________________________________
\section{Bulge Globular Clusters}

\subsection{Nature of our Sample}
Since the pioneering work of \citet{shapley18}, it has been recognized that the GCs of our Galaxy have a strong central concentration. Indeed, the density may  increase within a few kpc of the Galactic center, suggesting possibly distinct outer (halo) and inner (bulge) groups.

\citet{zinn85} first posited the existence of separate halo and disk populations of GCs based on their distinct spatial and metallicity distributions.  From existing information on metallicity,
scale height, and rotational velocities available at that time,
\citet{armandroff89} (and references therein) interpreted a sample
of low Galactic latitude metal-rich GCs as belonging to a disk system. However, \citet{minniti95}, from metallicity and kinematics of GCs in the central 3 kpc, suggested that such GCs instead constitute a
bulge population. This was corroborated by \citet{cote99}, who used spectroscopic metallicities and RVs for GCs within 4kpc of the Galactic center. The issue of whether there is a single bulge/(thick)disk population or they are distinct has been discussed for several decades (see e.g. \citealp{harris01}).

As recently as a few years ago, the review of \citet{bica+16} left this question still somewhat open. They defined BGCs as those with $R_{GC}<3kpc$ and [Fe/H]$\geq-1.5$ and found 43 such GCs in the H10 catalog. However, they also realized that these limits were somewhat arbitrary and admitted the possibility of such exceptions as halo intruders, with low metallicities which are at the moment simply passing thru the bulge near their perigalacticon  \citep[e.g. VVV\,CL001, ][]{Fernandez-Trincado+21}, or metal-rich GCs lying beyond the radial limit, which overlap with Armandroff's disk GCs but which did not have existing space velocities available at the time, prohibiting more precise characterization.

With the advent of the exquisite proper motions provided by Gaia, our ability to characterize GCs has been revolutionized by adding the powerful dimension of kinematics/dynamics. In fact, we now realize, in large part thanks to Gaia, that GCs can not only be classified as halo, bulge or disk but that indeed we can associate them with either an in situ or ex situ (accreted) origin, and often can also identify the latter with a particular accreted progenitor. A number of papers have carried out dynamical classifications of as many GCs as possible  (e.g. Massari et al. 2019). Two of the most recent of these are \citet{perez-villegas+20} and \citet{callingham+22}. Although Gaia is not as effective in deriving proper motions for heavily reddened, crowded BGCs, excellent data still exists for almost all of them, allowing rather definitive classification. What we have learned from this exercise is that most if not all halo GCs have been accreted and that there are indeed separate bulge and thick disk GC populations which were both born in situ. 

Looking at our sample of BGCs in \citet{perez-villegas+20}, we find that all but 3 are classified as bulge/bar clusters, while Terzan 12 and Ton 2 are deemed thick disk clusters and Terzan 10 is denoted as an inner halo cluster.
\citet{callingham+22} make very similar classifications, with all of our sample being assigned to the bulge
except BH 261, Terzan 10 and 12 and Ton 2, which are assigned to the Kraken progenitor.
These classifications also agree very well with those of \citet{massari+19}, with the exception that NGC 6401 was also listed as a Low Energy cluster, as was Ton 2, and Terzan 10 was associated with the Gaia-Enceladus-Sausage progenitor.
We note that Terzan 10 is the lowest metallicity cluster in our sample as well, so that its extraBulge and extraGalactic origin are not unexpected (see below).  However, it is important to also note that our RV is 270 km/s different from that of B19, so that one must be careful in using the correct RV when deriving orbits. In fact, \citet{ortolani+19} used a preliminary version  of our value to calculate Terzan 10's orbit and find that it is clearly a halo intruder, currently only passing through the bulge. We caution that our RV is also very different from that of B19 for Terzan 9. 
Thus, most of our clusters are indeed BGCs, with the notable exceptions of the most metal-poor and two of the four most metal-rich.

We finally note that the census of Galactic GCs, especially BGCs, is probably still quite incomplete. Near IR surveys like VVV/X have uncovered a large number of GC candidates in the bulge and adjacent disk in recent years \citep[e.g.][]{Minniti+10,moni-bidin+11,camargo19,palma+19,garro+22} and some of these, on close inspection  including the use of spectroscopy, turn out to indeed be previously unknown GCs  \citep[e.g.][]{dias+22b}. However, not all of these candidates turn out to be true GCs (e.g. \citealp{gran+19}, Minniti et al. 2021b, Geisler et al. 2021, G22)  and care must be taken to use all weapons at our disposal (central concentration and density, field correction, dereddening, RV, proper motion, CMD, abundances, presence of RR Lyrae, etc.) to correctly identify the true nature of such candidates in this very difficult region.

\subsection{The BGC metallicity distribution}
\label{sec:mdf}

Since the seminal work of \citet{zinn85}, it was known that a key distinguishing characteristic of halo and bulge/disk GCs was their metallicity distribution (MD). He showed that halo GCs are mostly metal-poor with a peak at [Fe/H]$\sim$ -1.6, whereas disk GCs are mostly metal-rich, peaking around -0.5, with an intermediate minimum near -0.8. Of course, the quantity and especially quality of metallicities available some 40 years ago was limited and very
crude compared to today's standards,
in particular for BGCs, yet the above basic impression of a unimodal metal-rich BGC system still persists. It is of some interest to revisit the BGC MD with modern data, including our own, and compare it with recent determinations of the bulge field star MD.

A relatively recent examination of BGCs was carried out by \citet{bica+16}, but using the H10 metallicities, which are still not optimum, as well as defining BGCs only by $R_{GC}$ and [Fe/H]. They found a bimodal MD for BGCs, with peaks around -0.5 and -1.1. The metal-rich
peak is of course the well-known classical \citet{zinn85} peak, but they demonstrated that the
metal-poor peak is perhaps dominant.

A more recent study is that of \citet{perez-villegas+20}, with metallicities taken from a wide variety of sources, but including their assessment of GC origin using Gaia DR2 proper motions. For their sample of 29 bonafide bulge/bar GCs, they uncovered the same two peaks as \citet{bica+16}, as well as a small group of only 3 GCs with even lower metallicities, ranging from -1.4 to -1.6. The reality of this possible low metallicity peak in the MD of genuine BGCs is of course of interest, particularly since such clusters, if indeed BGCs, could well include the oldest in situ GC in the Galaxy. Note that their sample does not include the halo interloper Terzan 10 noted above. More careful analysis including especially the best metallicities as well as RVs available is needed to definitively assess this issue.

\begin{figure}
    \centering
    \includegraphics[width=\columnwidth]{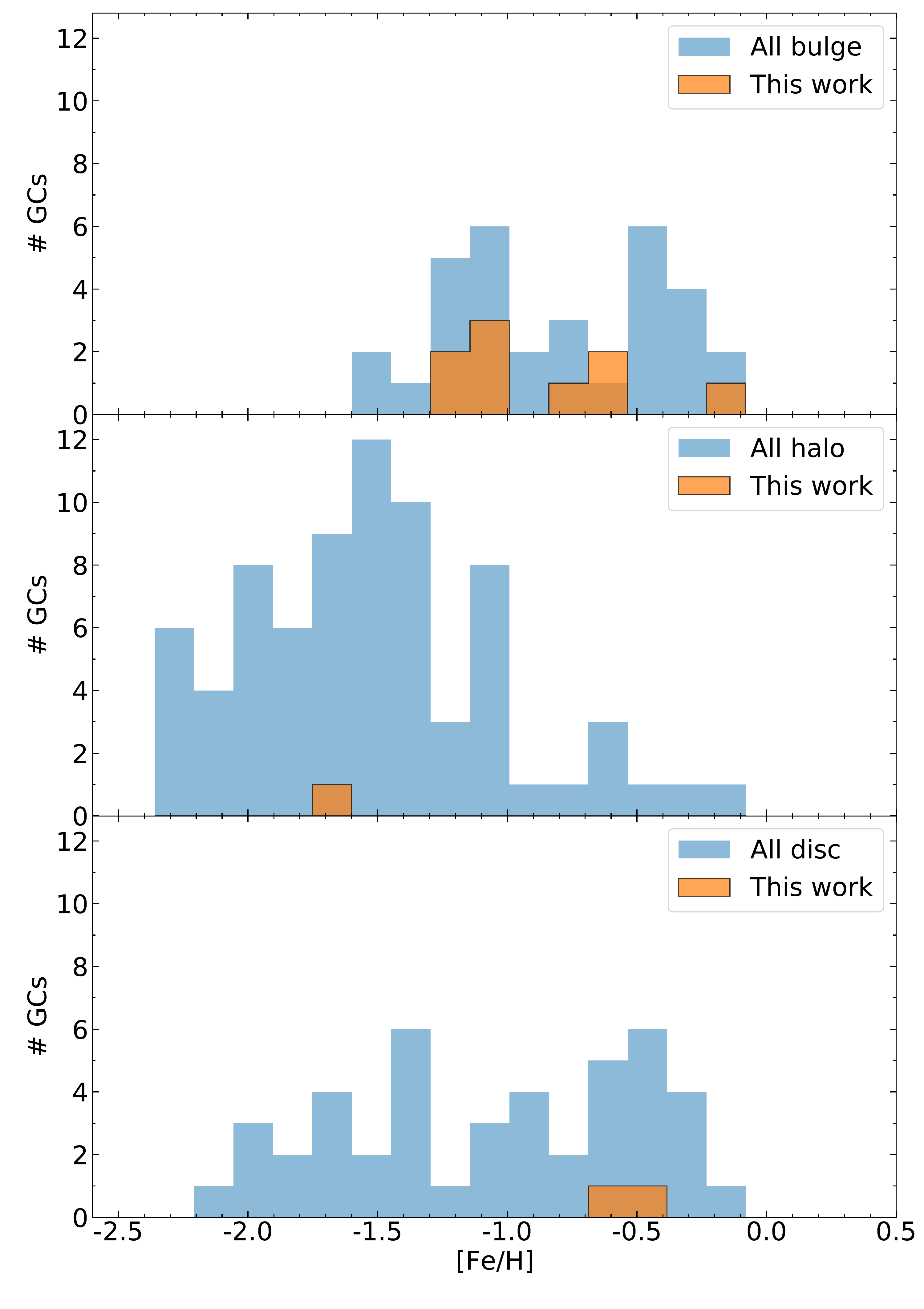}
    \caption{Metallicity distribution function of all Galactic GCs from the metallicity scale of \citet[][2019 version]{dias+16,dias+16b} shown as a blue histogram split by population following the same classification adopted in Figure \ref{fig:aitoff}. The orange histogram represents the respective GC samples analysed in this work adopting the metallicities derived here applying the  V18 % DP 
    scale as shown in Table \ref{t:results}.}
    \label{fig:mdf}
\end{figure}

We plot our results in Figure \ref{fig:mdf}, where we compare our values to those from the Dias et al. catalog for different GC classes. Our bonafide BGC sample is distributed over and between the two main BGC metallicity peaks. Our data does not strongly support a simple bimodal distribution with the above peak values, but 
clearly our sample is too small to draw any definitive conclusions from.
Again, better metallicities for the largest possible sample are required to clarify the nature of the BGC metallicity distribution. Our halo GC and 2 disk GCs fall nicely within the MD of 
their class.
Note again that our lowest metallicity cluster, Terzan 10, is now considered to be an inner halo cluster and not a BGC. Our most metal-poor bonafide BGC is BH 261 at [Fe/H] = -1.21. 

Are there any true BGCs with metallicities lower than $[Fe/H] \sim$ -1.25? Our small sample does not contain any. However, the CAPOS study of G21 does find three such clusters, Terzan 4, 9 and HP 1, with [Fe/H] from -1.2 to -1.4, all classified as bulge by \citet{perez-villegas+20} and Callingham et al. (2022). As discussed above, CAPOS metallicities are substantially  lower
in the mean than our CaT values for clusters in common, as is the case for Terzan 9. There is strong interest in determining reliable metallicities and ages for such metal-poor BGCs, in particular those with a blue horizontal branch, as they are excellent candidates for the oldest native GCs of the Milky Way \citep{lee+94,barbuy+06,dias+16,barbuy+18_6558}, since they were born in situ. Despite the fact that such clusters are more metal-rich than the peak of the halo MD, they could 
indeed be older than their lower metallicity halo couterparts given the expected more rapid chemical evolution in the deeper potential well of the proto-Galaxy as opposed to the shallower wells of much lower mass progenitors that generated the accreted halo GCs (Cescutti et al. 2008).

We also compare our BGC MD with that of bulge field stars. Probably the best recent bulge field star MD was derived by  \citet{rojas-Arriagada+20} from APOGEE spectra. They compiled a total of $\sim $13000 bulge stars and find
strong evidence for trimodality, with peaks at [Fe=H] = +0.32,
-0.17 and -0.66. These peaks maintain their value but their
relative strengths vary as a function of Galactic latitude. The
fraction of stars below -1 is very small, in contradistinction to
our sample. It is likely that the most metal-poor field-star peak and metal-rich GC peak have similar origins. However, it is unclear why the field and GC MDs are otherwise quite distinct. 
One possible explanation could be quite different age distributions (i.e. different formation epochs).  Although GCs are all “old”, i.e. $>10$ Gyr or so, we don’t have very accurate ages for the field stars, which could be somewhat younger and thus more metal-rich. Note that bulge RR Lyrae stars  peak around [Fe/H]=-1.0 \citep{dekany+13}.
Clearly, further improvement of both  the quality and number of BGC metallicities is required to help address such puzzles. 

\section{Conclusions} 

We have obtained low resolution spectra of the CaT lines in a total of about 540 red giants in the vicinity of 13 bulge GCs using the FORS2 instrument on the Very Large Telescope. The targeted clusters were those which did not have spectra of individual stars at the time of our observations because of high reddening and therefore had rather poor existing estimates of RVs and especially metallicities.
We measured the wavelengths and equivalent widths of the CaT lines and derived RVs and metallicities using standard 
procedures. An extensive membership assessment involving position in the cluster and CMD, RV, metallicity and proper motion insured very high membership probabilities for our final sample. Unfortunately, one of our clusters, VVV CL002, turned out to not to have any members among our observed stars. We derive mean cluster RV values with a mean standard error of the mean of 3 km/s, and mean metallicities to 0.05 dex for an average of 11 members per cluster for the remaining sample.  

Next, we compared our mean RVs and metallicities with previous literature values for each cluster, focusing on determinations from five recent, internally homogeneous catalogues. Overall
agreement with the published RV values is generally good, but suggests our values are about 3 km/s lower on average. As for
[Fe/H], our mean values tend to  agree with V18
and to a lesser extent with D19, which are on the same metallicity scale.
Particularly puzzling is a mean offset of  about 0.2 dex to
higher metallicities for our results compared to the recent high resolution study of G22.

We then discuss the nature of our GCs, finding that almost all of them are indeed bonafide BGCs. However, the most metal-poor cluster, Terzan 10, is likely a halo intruder, as first noted by Ortolani et al. (2019), while two of the most metal-rich are likely thick disk GCs. Finally, we examine the metallicity distribution of BGCs, comparing them to both halo and disk GCs, as well as to bulge disk stars. Our BGCs  roughly follow the MD of other BGCs with metallicities on the same scale, including 5 clusters with [Fe/H]$\lesssim-1.0$. 
The possibility of a small, even more metal-poor group  ([Fe/H]$\lesssim-1.25$) is currently unclear, and may only include further halo interlopers. The metal-rich peak coincides with the metal-poorest peak of the trimodal bulge field star distribution, leaving the more metal-poor BGCs with very few field star counterparts except for the RR Lyrae.

\begin{acknowledgements}
D.G. gratefully acknowledges support from the ANID BASAL project ACE210002.
D.G. also acknowledges financial support from the Direcci\'on de Investigaci\'on y Desarrollo de
la Universidad de La Serena through the Programa de Incentivo a la Investigaci\'on de
Acad\'emicos (PIA-DIDULS). S.V. and D.G. gratefully acknowledge the support provided by Fondecyt reg. 1220264. S.V. also acknowledges the the support provided by ANID BASAL projects ACE210002 and  FB210003.
This research was partially supported by the Argentinian institutions CONICET, SECYT (Universidad Nacional de Córdoba) and Agencia Nacional de Promoción Científica y Tecnológica (ANPCyT). B.D. acknowledges support by ANID-FONDECYT iniciación grant No. 11221366. 
D.M. gratefully acknowledges support by the ANID BASAL projects ACE210002 and FB210003 and by Fondecyt Project No. 1220724. 
We appreciate helpful comments raised by the referee. \\
Based on observations collected at the European Southern Observatory under ESO programmes 089.D-0392 and 091.D-0389(A).
This research has made use of the services of the ESO Science Archive Facility.\\
This work has made use of data from the European Space Agency (ESA) mission
{\it Gaia} (\url{https://www.cosmos.esa.int/gaia}), processed by the {\it Gaia}
Data Processing and Analysis Consortium (DPAC,
\url{https://www.cosmos.esa.int/web/gaia/dpac/consortium}). Funding for the DPAC
has been provided by national institutions, in particular the institutions
participating in the {\it Gaia} Multilateral Agreement.\\

\end{acknowledgements}

%%%%%%%%%%%%%%%%%%%%%%%%%%%%%%%%%%%%%%
%___________________________________________________________
   \bibliographystyle{aa} % style aa.bst
   \bibliography{bibliography} % your references Yourfile.bib

%%%%%%%%%%%%%%%%%%%%%%%%%%%%%%%%%%%%%%
%\appendix

\end{document}